\documentclass[12pt]{article}

\usepackage{amsmath, amssymb}
\usepackage{graphicx}
\usepackage{booktabs}
\usepackage{geometry}
\usepackage{hyperref}
\usepackage{chngcntr}
\usepackage{caption}
\usepackage{subcaption}
\usepackage{booktabs}
\usepackage{multirow}
\usepackage{float}
\usepackage{xurl}
\numberwithin{figure}{section}
\title{MAT-MPNN: A Mobility-Aware Transformer-MPNN Model for Dynamic Spatiotemporal Prediction of HIV Diagnoses in California, Florida, and New England}
\author{
Zhaoxuan Wang$^{1,\dagger}$, Weichen Kang$^{1,\dagger}$, Yutian Han$^{1,\dagger}$, \\ Lingyuan Zhao$^{1,\dagger}$, Prof. Bo Li$^{1,*}$ \\
\small $^{1}$ Department of Statistics, Washington University in St.Louis, Saint Louis, United States \\
\small $\dagger$ These authors contributed equally. \\
\small $^{*}$ Corresponding author: \texttt{bol@wustl.edu}
}
\date{}

\begin{document}
\maketitle

\begin{abstract}
Human Immunodeficiency Virus (HIV) has posed a major global health challenge for decades, and forecasting HIV diagnoses continues to be a critical area of research. However, capturing the complex spatial and temporal dependencies of HIV transmission remains challenging. Conventional Message Passing Neural Network (MPNN) models rely on a fixed binary adjacency matrix that only encodes geographic adjacency, which is unable to represent interactions between non-contiguous counties. Our study proposes a deep learning architecture Mobility-Aware Transformer–Message Passing Neural Network (MAT-MPNN) framework to predict county-level HIV diagnosis rates across California, Florida, and the New England region\footnote{New England includes seven U.S. states: Connecticut, Delaware, Maryland, Massachusetts, New Jersey, New York, and Pennsylvania.}. The model combines temporal features extracted by a Transformer encoder with spatial relationships captured through a Mobility Graph Generator (MGG). The MGG improves conventional adjacency matrices by combining geographic and demographic information. Compared with the best-performing hybrid baseline, the Transformer MPNN model, MAT-MPNN reduced the Mean Squared Prediction Error (MSPE) by $27.9\%$ in Florida, $39.1\%$ in California, and $12.5\%$ in New England, and improved the Predictive Model Choice Criterion (PMCC) by $7.7\%$, $3.5\%$, and $3.9\%$, respectively. MAT-MPNN also achieved better results than the Spatially Varying Auto-Regressive (SVAR) model~\cite{shand2018svar} in Florida and New England, with comparable performance in California. These results demonstrate that applying mobility-aware dynamic spatial structures substantially enhances predictive accuracy and calibration in spatiotemporal epidemiological prediction.
\end{abstract}


\section{Introduction}

Human Immunodeficiency Virus (HIV) has always been one of the most critical global public health challenges. In recent estimates, millions of people in the world living with HIV, with a large number of new infections each year. In 2024, an estimated 1.3 million people acquired HIV globally~\cite{unaids2025}. This makes data-driven prediction for prevention extremely important.

Previous studies, such as Li et al.~\cite{shand2018svar}, analyzed county-level HIV diagnoses rate across three regions in United States: California, Florida, and New England, using data available only through 2014. Their work provided valuable insights into spatiotemporal modeling of epidemics such as HIV. However, the limited temporal coverage restricts its ability to reflect such rich dynamics. This gap motivates our study to extend HIV prediction to more recent years and to explore modern spatiotemporal deep leaning methods for capturing the mobility patterns of HIV diagnoses.

Traditional approaches to spatiotemporal modeling are trying to capture complex dependencies across both time and space using deep learning architectures designed for sequential and relational data. Recurrent Neural Networks (RNNs)~\cite{elman1990finding}, particularly Long Short-Term Memory (LSTM) networks, have been widely applied to model temporal dependencies~\cite{hochreiter1997lstm,nguyen2023covid}. However, their sequential processing and reliance on gated mechanisms (e.g., the forget gate) limit their ability to capture long-range temporal relationships. Transformer~\cite{vaswani2017attention} overcome these constraints by applying self-attention to learn global temporal dependencies. On the spatial side, Graph Neural Networks (GNNs)~\cite{kipf2017gcn,wu2021gnnsurvey} are powerful tools for modeling geographically relational data. They represent entities as nodes and the relationships between them as edges within a graph structure. Among them, Graph Convolutional Networks (GCNs)~\cite{kipf2017gcn} are the most fundamental form for conventional GNNs. The graph convolution operations aggregate information from neighboring nodes to learn spatial representations. Building upon the foundation of GNNs, Message Passing Neural Networks (MPNNs)~\cite{gilmer2017mpnn,kapil2022covid} and Graph Attention Networks (GATs)~\cite{velickovic2018gat} have shown strong abilities capturing spatial correlations through graph structures. However, these architectures are often limited to a single axis: sequence models like LSTMs and Transformers capture temporal evolution per location without explicit spatial coupling, while graph-based models like GCNs, MPNNs, and GATs focus on spatial structure within static snapshots without explicit temporal dynamics. To overcome these limitations, hybrid spatiotemporal models have been developed to jointly learn spatial and temporal dependencies. For instance, Kapoor et al.~\cite{kapoor2020examining} proposed a hybrid graph–temporal architecture that integrates human mobility networks with graph neural networks to improve COVID-19 forecasting accuracy. Similarly, Zhu et al.~\cite{zhu2024modeling} developed a Graph Attention–based Spatial–Temporal (GAST) hybrid model that jointly learns temporal dynamics and spatial dependencies. This makes dynamic attention weights to represent heterogeneous inter-regional influence. Moreover, Scott et al.~\cite{nandy2025socio} introduced a hybrid dynamic graph framework capable of learning adaptive edge structures as part of the temporal modeling process. In conclusion, these studies demonstrate that hybrid spatiotemporal models, which integrate temporal sequence learning with spatial graph structures, can significantly enhance predictive performance. These works form the method foundation for our approach. Nevertheless, as Scott et al.~\cite{nandy2025socio} emphasized, “geography is important but not all-important.” Their model shows that geographical distant regions can still share strong socio-demographic similarities, challenges the traditional first law of geography\footnote{Tobler’s First Law of Geography states that “everything is related to everything else, but near things are more related than distant things”.}~\cite{tobler1970law}. This motivates us to capture mobility-driven and demographic relationships beyond simple geographic adjacency when modeling disease spread.

Building upon these perspectives, our study addresses a key limitation of existing hybrid spatiotemporal models, which use fixed geography-based adjacency matrices that cannot capture relationships between distant but correlated regions. To overcome this, we propose a Mobility-Aware Transformer–Message Passing Neural Network (MAT-MPNN) framework that integrates temporal and spatial learning through an adaptive graph construction mechanism. The proposed method introduces the Mobility Graph Generator (MGG) to dynamically optimize the adjacency structure based on both mobility and demographic similarity, which allows the model to more accurately reflect real world inter-county interactions. Specifically, \\(1) at the theoretical level, our design is inspired by Kalofolias~\cite{kalofolias2016learn}, who demonstrated that a graph structure can be learned directly from smooth node signals by optimizing edge weights so that connected nodes exhibit similar feature values; \\(2) at the architectural level, we integrate a Transformer encoder for long range temporal dependencies with an MPNN layer for spatial message passing, achieving a coherent representation of spatiotemporal dynamics; \\and (3) at the application level, we implement this framework to predict county-level HIV diagnosis rates, and the result shows an improved predictive accuracy across the three regions. In conclusion, this innovation creates a flexible and mobility-aware spatiotemporal framework that better captures relationships beyond simple geographic boundaries.

\section{Data}

\subsection{Overview and Study Regions}

We obtained county-level annual HIV diagnosis rates from 2008 to 2022 from the Centers for Disease Control and Prevention (CDC) AtlasPlus platform and AIDSVu.org~\cite{cdc_aidsvu}. HIV diagnosis rates are reported as the number of new diagnoses per 100{,}000 population for each U.S. county. These data originate from the CDC’s \textit{National HIV Surveillance System (NHSS)} and are publicly disseminated through the \textit{AIDSVu County-Level HIV Diagnoses and Prevalence Dataset (2008--2022)}, developed by Emory University’s Rollins School of Public Health in partnership with the CDC Division of HIV Prevention. To ensure confidentiality, data are suppressed under several conditions~\cite{shand2018svar}: \\(a) when a county reports fewer than five cases or has a population smaller than 100, \\(b) when state health departments opt not to release data to AIDSVu due to data-use agreements with the CDC, \\or (c) when county-level data are unavailable, as in Alaska, the District of Columbia, or Puerto Rico.

In the \textit{AIDSVu County-Level HIV Diagnoses and Prevalence Dataset (2008--2022)}, suppressed or missing values are represented by numerical codes. Specifically, the values $-1$, $-2$, $-4$, $-8$, and $-9$ indicate different suppression or data unavailable conditions. $-1$ means that the data is not shown to protect privacy due to small case counts or small population sizes. $-2$ indicates that the data were not released to AIDSVu because state health departments, following their data re-release agreements with the CDC, requested suppression of data for counties below certain population thresholds. These agreements are periodically updated, which may cause annual differences in the availability of county-level data. $-4$ denotes that data is not available at the county-level. $-8$ corresponds to undefined values, for example when rate or proportion cannot be calculated because the denominator is $0$. Finally, $-9$ represents data that are unavailable. These suppression codes are standardized across AIDSVu data products to ensure individual privacy. All county and year observations containing any of these suppression codes ($-1$, $-2$, $-4$, $-8$, or $-9$) were excluded from the analysis to ensure data completeness and consistency among predictors. 

Because of these suppression rules and the relative sparsity of HIV in certain areas, only a subset of county–year observations is available. Therefore, we focused on three regions with relatively complete records and high HIV prevalence: California (40 counties), Florida (61 counties), and the New England region (Connecticut, Delaware, Maryland, Massachusetts, New Jersey, New York, and Pennsylvania, 149 counties), which collectively comprise 250 counties with an average of more than 70\% non-missing annual observations~\cite{shand2018svar}.

\subsection{Covariates}

Beyond HIV surveillance data, we incorporated additional socioeconomic, demographic, and structural covariates from multiple public sources. We collected data from the U.S. Census Bureau’s \textit{Small Area Health Insurance Estimates (SAHIE)}~\cite{census_sahie} to provide annual county-level indicators of health insurance coverage, income, and poverty between 2009 and 2022. Since data for 2008 were unavailable, we treated the 2008 HIV diagnosis rate as a covariate to predict 2009 outcomes. More generally, the previous year’s diagnosis rate was treated as a covariate in all years. We also obtained general demographic data, including population size, sex, and age composition, from the \textit{American Community Survey (ACS) 5-Year Estimates} (Table B15002: Sex by Educational Attainment for the Population 25 Years and Over)~\cite{census_acs_b15002}. Moreover, we obtained county-level incarceration rate data from the \textit{Vera Institute of Justice Incarceration Trends Dataset}~\cite{vera_incarceration}. Additionally, we extracted county-level hospital location data from the \textit{Homeland Infrastructure Foundation-Level Data (HIFLD)}~\cite{hifld_hospitals}, which provides the spatial distribution of hospitals across the United States.

We obtained geospatial information used to align and visualize counties from the \textit{U.S. Census Bureau’s Cartographic Boundary Shapefiles (2022)}~\cite{census2022shapefile}, which include both county and state boundary geometries for matching geographic identifiers (GEOIDs). Moreover, We extract county-level transportation connectivity features from the \textit{National Transportation Atlas Database (NTAD)}~\cite{ntad_transportation} from the U.S. Bureau of Transportation Statistics, which contains data on Amtrak station locations. This data is used to compute the number of train stations per county. All data sources were merged using county-level GEOIDs.

In addition to these covariates, we expanded the socioeconomic and demographic feature set using the \textit{tidycensus} R package~\cite{tidycensus}, which provides direct access to the U.S. Census Bureau’s \textit{American Community Survey (ACS) 5-Year Estimates}~\cite{census_acs}. We retrieved $473$ county-level variables across multiple ACS tables, including \texttt{B23001} (Sex by Age by Employment Status), \texttt{B27001} (Health Insurance Coverage Status by Sex and Age), \texttt{B18101} (Disability Status by Age and Sex), \texttt{B16001} (Language Spoken at Home), \texttt{B05002} (Place of Birth by Citizenship Status), and other tables related to housing characteristics, transportation behavior, broadband access, veteran status, race, and household composition. These tables were selected to capture social and economic dimensions that may influence HIV transmission dynamics and healthcare access.

\subsection{Data Alignment and Integration}

We downloaded the ACS data for all counties and years corresponding to the HIV dataset (2008–2022) and aligned using Federal Information Processing Standards (FIPS) county codes (\texttt{GEOID}) and survey year. Each ACS table was retrieved by the \texttt{get\_acs()} function in \textit{tidycensus} R package and merged into the analysis dataset by aligning on county–year pairs.

\begin{figure}[htbp]
    \centering
    \numberwithin{figure}{subsection}
    \includegraphics[width=0.8\textwidth]{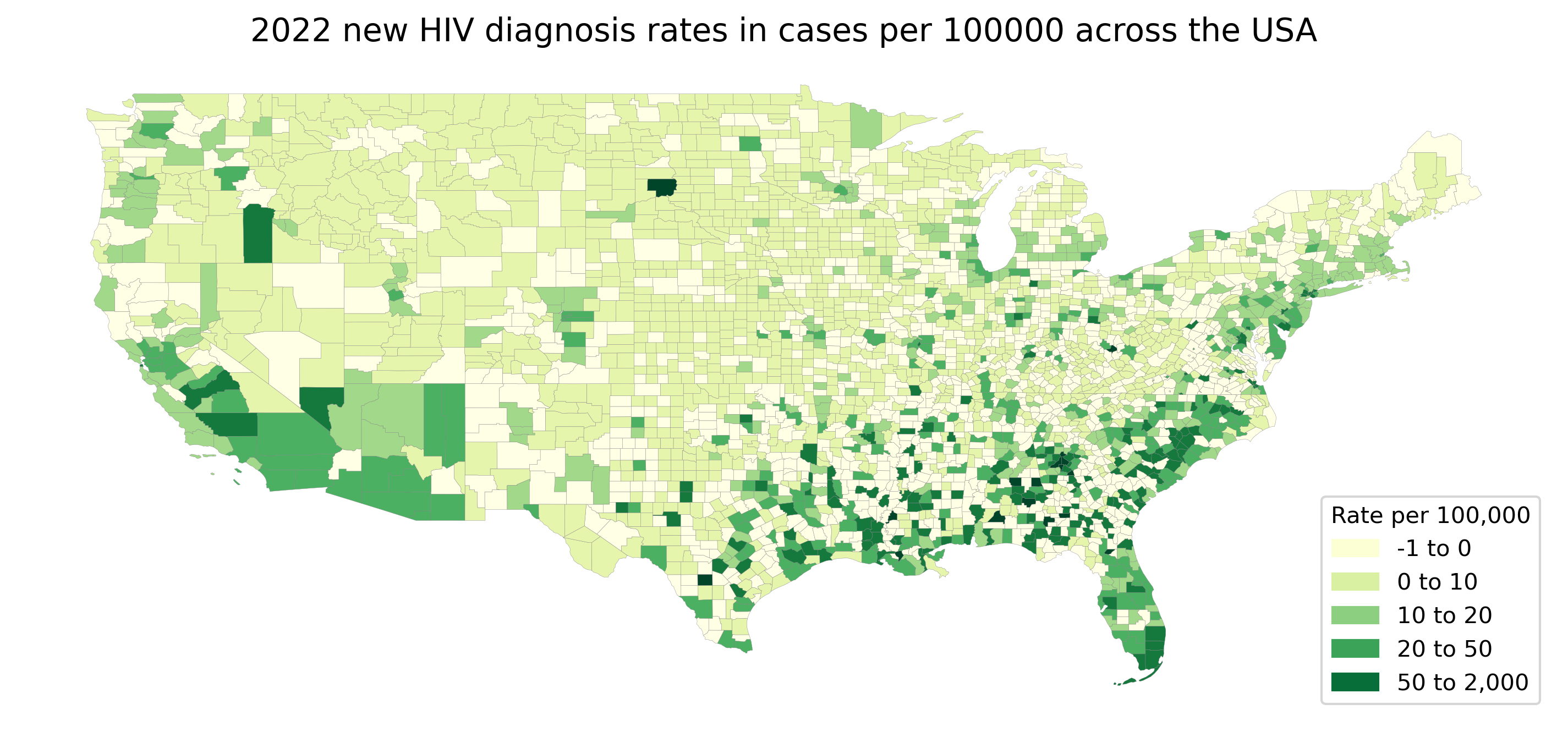}
    \caption{County level 2022 new HIV diagnosis rates in cases per 100{,}000 across the USA}
    \label{fig:overall_map}
\end{figure}

The dataset covers 15 years (2008–2022) and includes 498 county-level variables from different public sources. These variables describe demographic, economic, mobility, infrastructure, and health-related factors. The combined dataset serves as the basis for building the time-varying mobility-aware adjacency matrix used in the MAT-MPNN model.

\section{Methodology} 

Our goal is to model how HIV rates evolve within each county over time and how they spread across counties through spatial relationships. So we will capture both the temporal evolution of HIV diagnoses and the spatial diffusion of the epidemic across counties.
To achieve this, we design a hybrid framework that integrates three complementary components:
(1) a Transformer encoder for learning long-range temporal patterns;
(2) a Mobility Graph Generator (MGG) that constructs dynamic inter-county connections; and
(3) a temporal Message Passing Neural Network (MPNN) that performs graph-based spatial aggregation.

\subsection{Model Overview}

\begin{figure}[H]
    \centering
    \numberwithin{figure}{subsection}
    \includegraphics[width=0.9\textwidth]{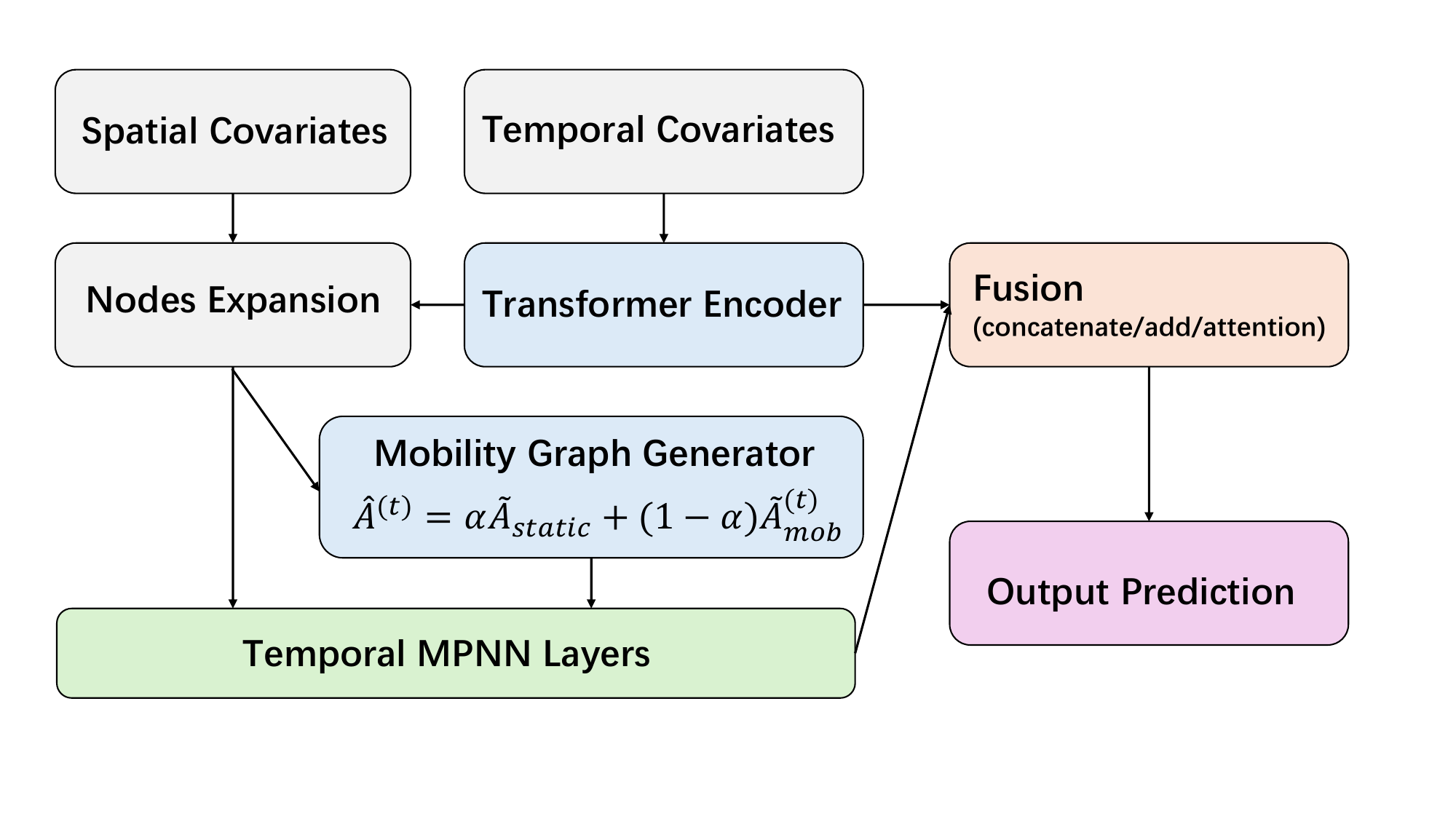}
    \caption{Overall architecture of the proposed MAT–MPNN framework. The model integrates temporal learning via a Transformer encoder, adaptive graph construction via the Mobility Graph Generator (MGG), and spatial message passing via the MPNN to jointly capture temporal evolution and spatial diffusion of HIV diagnoses.}
    \label{fig:mat_mpnn_architecture}
\end{figure}

Fig.~\ref{fig:mat_mpnn_architecture} presents the overall architecture of Mobility-Aware Transformer–MPNN (MAT–MPNN) framework. The model captures the temporal evolution and spatial diffusion of HIV diagnoses across U.S. counties. Temporal covariates are first processed through a Transformer encoder to learn long-range temporal dependencies. In parallel, static spatial covariates, including geographic and infrastructure features, are expanded into node representations to align with the temporal dimension. The Mobility Graph Generator (MGG) then constructs a dynamic adjacency matrix, which combines static geographic connectivity with mobility inter-county similarity. This adaptive graph serves as a dynamic map that allows the Temporal MPNN layers to share information between geographically and mobility-connected counties. The outputs from the Transformer and MPNN components are fused through concatenation or attention mechanisms to form a unified latent representation, from which the model predicts county-level HIV diagnosis rates for each year.

\subsection{Input Decomposition and Linear Projection}

The dataset is separated into time-varying and time-invariant covariates. Time-varying covariates include $495$ variables that change annually across counties from 2008 to 2022. Time-invariant covariates describe static geographic attributes that remain constant across years, including county land area, hospital count, and train station count.

Let
\begin{align}
    X_{\text{temporal}}\in \mathbb{R}^{B\times T\times F_t}, \quad 
    X_{\text{spatial}}\in \mathbb{R}^{B\times F_s}, 
    \label{eq:input_decomposition}
\end{align}
where $B$ is the batch size, $T$ is the number of years, and $F_t$, $F_s$ denote the feature dimensions of the temporal and spatial covariates, respectively. This separates time-varying and time-invariant information to ensure that temporal and spatial features are represented in a comparable latent space before integration.

Then apply separate linear projections:

\begin{align}
    Z = X_{\text{temporal}} W_t^{\top} + b_t, 
    \label{eq:temporal_projection} \\
    S = X_{\text{spatial}} W_s^{\top} + b_s, 
    \label{eq:spatial_projection}
\end{align}
where $Z\in\mathbb{R}^{B\times T\times d}$ and $S\in\mathbb{R}^{B\times d}$ are the projected embeddings, and $W_t$, $W_s$ are learnable weight matrices. 

The linear projections transform both the time-varying and time-invariant covariates into low-dimensional representations that capture the most informative patterns for subsequent temporal and spatial modeling.

\subsection{Temporal Encoding}

The projected temporal embeddings $Z$ are fed into a Transformer encoder~\cite{vaswani2017attention} to model temporal dependencies:

\begin{align}
    H = \text{Transformer}(Z).
    \label{eq:transformer_encoding}
\end{align}

The Transformer encoder applies multi-head self-attention, where each time step attends to all others:

\begin{align}
    \text{Attention}(Q, K, V) = \text{softmax}\!\left(\frac{QK^{\top}}{\sqrt{d_k}}\right)V, 
    \label{eq:self_attention}
\end{align}
where $Q$, $K$, and $V$ denote the query, key, and value matrices obtained through learned linear transformations of the input $Z$. The Transformer encoder enables the model to identify which past years are most relevant to current conditions, resulting to capture cumulative temporal effects. Each county’s time series is processed independently, allowing the encoder to learn a dynamic temporal representation $H_i$ for each county.

To align feature dimensions for subsequent spatial interaction, the temporal embeddings $H$ are duplicated across all counties $N$, and the static spatial embeddings $S$ are expanded across all time steps $T$ and counties $N$. This produces two tensors with identical dimensions, which are then combined element-wise to obtain the unified representation:
\begin{align}
    X = \tilde{H} + \tilde{S}.
    \label{eq:combine_embeddings}
\end{align}

\begin{figure}[H]
    \centering
    \includegraphics[width=0.9\textwidth]{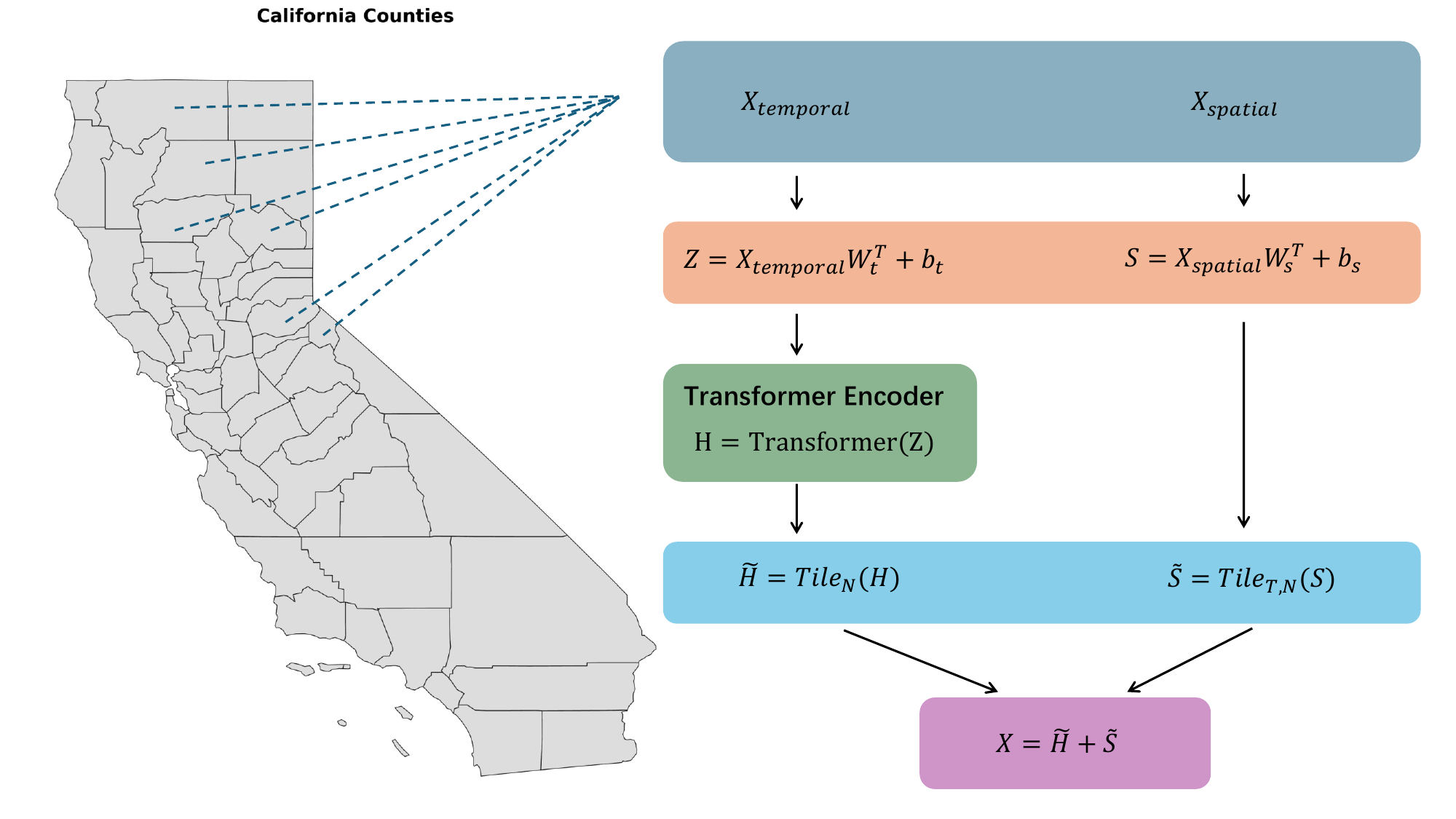}
    \caption{Temporal–spatial encoding pipeline using California counties as an example. 
    Each county provides temporal and spatial covariates, which are linearly projected 
    and processed through a Transformer encoder to produce temporally contextualized embeddings $H$. 
    Both $H$ and $S$ are duplicated and combined element-wise to form the unified representation 
    $X = \tilde{H} + \tilde{S}$.}
    \label{fig:step12}
\end{figure}

Fig.~\ref{fig:step12} shows how county-level features from California are separated into temporal and spatial components, transformed through linear projections, and processed by the Transformer encoder to produce a unified representation. The temporal embedding $H$ is then tiled across all counties $N$, and the spatial embedding $S$ is tiled across all time steps $T$ and all counties $N$ to align dimensions for subsequent spatial interaction.

\subsection{Mobility Graph Generator}

At each time step $t$, a dynamic adjacency matrix $\hat{A}^{(t)}$ is generated to represent inter-county interactions.

Each county feature $X_{t,i}$ is first re-encoded into a latent representation:
\begin{align}
    U_{t,i} = \phi(X_{t,i}) = \sigma(WX_{t,i} + b),
    \label{eq:county_encoding}
\end{align}
where $W\in \mathbb{R}^{d \times d_m}$ and $b\in \mathbb{R}^{d_m}$ are learnable parameters, and $\phi(\cdot)$ denotes a nonlinear activation that maps counties into an embedding space where similarities can be effectively compared.

\begin{align}
    e_{ij}^{(t)} = [U_{t,i} \| U_{t,j}] \in \mathbb{R}^{2d_m},
    \label{eq:concat_pair}
\end{align}
and an edge score is computed through a two-layer MLP:
\begin{align}
    s_{ij}^{(t)} = \text{MLP}(e_{ij}^{(t)}) 
    = W_2^{\top}\sigma(W_1 e_{ij}^{(t)} + b_1) + b_2,
    \quad s_{ij}^{(t)} \in \mathbb{R},
    \label{eq:mlp_edge}
\end{align}
where $\sigma$ denotes a nonlinear activation function, and $W_1$, $W_2$, $b_1$, and $b_2$ are learnable parameters. The concatenation enables the network to consider both counties jointly, and the MLP learns how similarity in socio-economic and health profiles translates into potential mobility or influence strength. 

After computing all pairwise edge scores $s_{ij}^{(t)}$, the raw mobility graph $A_{\text{mob}}^{(t)}$ is typically dense, as every county is connected to all others. To improve interpretability, top-$k$ sparsification is applied to each row of $A_{\text{mob}}^{(t)}$. Formally, for each county $i$ at time $t$, only the $k$ largest edge scores ($k=5$) are retained:
\begin{align}
    \tilde{s}_{ij}^{(t)} =
    \begin{cases}
        s_{ij}^{(t)}, & \text{if } j \in \mathcal{N}_i^{(k)} = \operatorname*{arg\,topk}_{j'} s_{ij'}^{(t)}, \\
        -\infty, & \text{otherwise.}
    \end{cases}
    \label{eq:topk_sparsify}
\end{align}
A row-wise softmax is then applied to obtain normalized connection weights, ensuring that each county’s outgoing weights sum to one:
\begin{align}
    A_{\text{mob}}^{(t)}(i,j) 
    = \text{softmax}(\tilde{s}_{ij}^{(t)}) 
    = \frac{\exp\!\left(\tilde{s}_{ij}^{(t)}/\tau\right)}
    {\sum_{j'}\exp\!\left(\tilde{s}_{ij'}^{(t)}/\tau\right)},
    \label{eq:softmax_norm}
\end{align}
where the temperature $\tau = 2$ smooths the distribution, preventing any county from dominating the connections.

A static adjacency matrix $A_{\text{static}}$ is constructed to encode fixed geographic relationships between counties, where the entry $A_{\text{static}}(i,j)$ is set to $1$ if counties $i$ and $j$ share a border, and $0$ otherwise. This matrix captures persistent geographic contiguity and serves as a stable baseline of spatial connectivity.

To compare the two graphs on the same scale, both are normalized using symmetric degree normalization with self-loops:
\begin{align}
    \tilde{A}_{\text{mob}}^{(t)} 
    = D_{\text{mob}}^{-\frac{1}{2}}(A_{\text{mob}}^{(t)} + I) D_{\text{mob}}^{-\frac{1}{2}}, 
    \quad
    \tilde{A}_{\text{static}} 
    = D_{\text{static}}^{-\frac{1}{2}}(A_{\text{static}} + I) D_{\text{static}}^{-\frac{1}{2}},
    \label{eq:adj_norm}
\end{align}
where $D_{\text{mob}}$ and $D_{\text{static}}$ denote the degree matrices of $A_{\text{mob}}^{(t)}$ and $A_{\text{static}}$, respectively. 

Finally, the normalized static and dynamic adjacency matrices are combined to form the final time-varying adjacency:
\begin{align}
    \hat{A}^{(t)} 
    = \alpha \tilde{A}_{\text{static}} 
    + (1 - \alpha)\tilde{A}_{\text{mob}}^{(t)},
    \label{eq:combined_adj}
\end{align}
where $\alpha \in [0, 1]$ is a learnable coefficient that balances the static geographic structure with dynamic mobility interactions. The blended adjacency $\hat{A}^{(t)}$ represents a unified graph that captures both persistent geographic and time-varying mobility relationships. 

\begin{figure}[H]
    \centering
    \numberwithin{figure}{subsection}
    \includegraphics[width=0.9\textwidth]{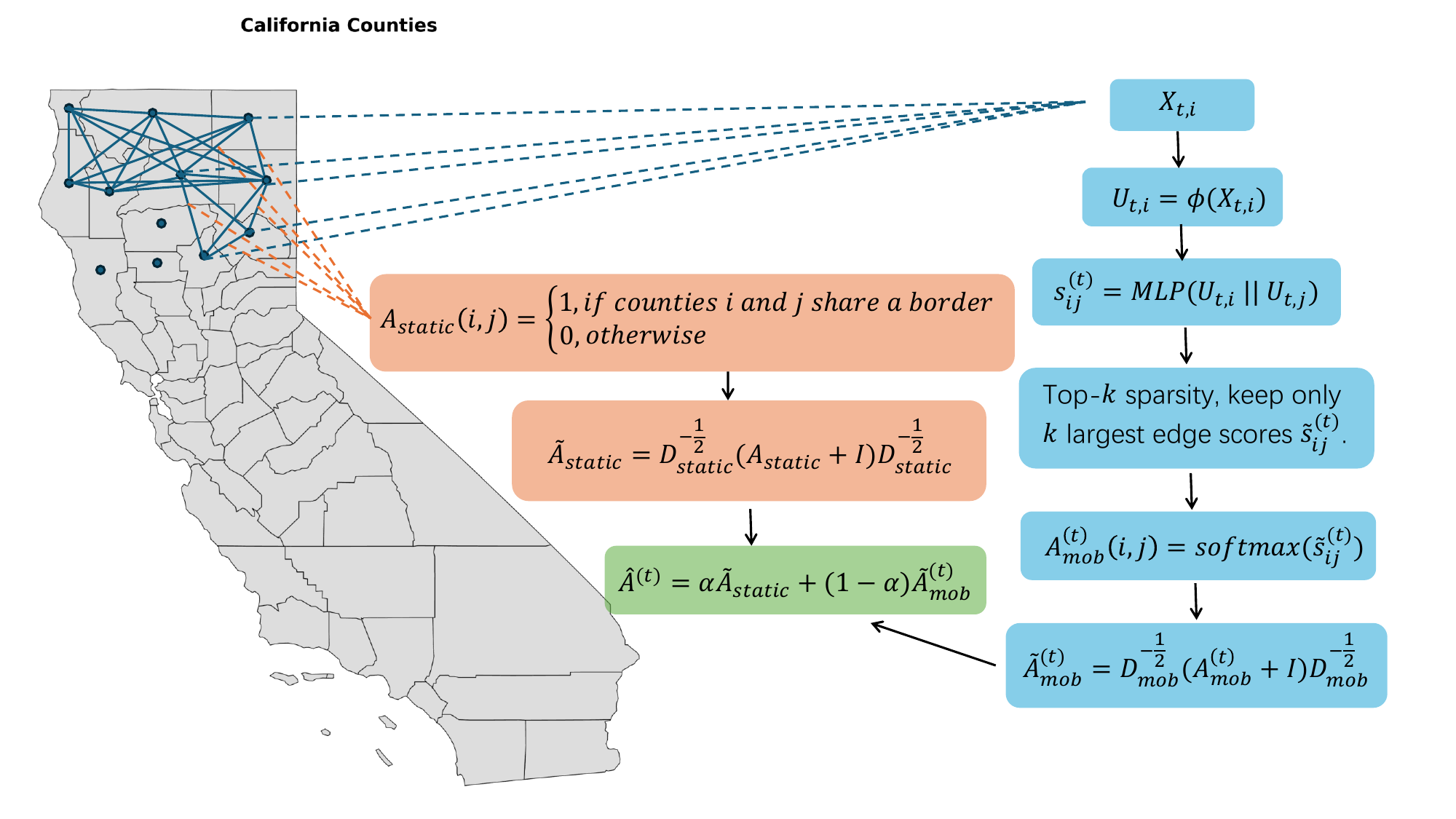}
    \caption{The MGG constructs time-varying adjacency matrices by encoding county embeddings, then score and choose the most important edges, normalizing their connection strengths, and integrating them with static geographic borders to capture both dynamic and static spatial dependencies.}
    \label{fig:step3}
\end{figure}

Fig.~\ref{fig:step3} shows how MGG combines learned mobility connections with static geographic borders to produce dynamic adjacency matrices for each time step.

\subsection{Spatial Message Passing}

After constructing the blended adjacency matrix $\hat{A}^{(t)}$, spatial message passing is applied across counties to propagate information through the dynamic mobility network. For each year $t$ and each message-passing layer $l$, the aggregated neighborhood information and node updates are computed as:
\begin{align}
    M_t^{(l)} = \hat{A}^{(t)} X_t^{(l)}, \quad 
    Y_t[i] = W_2 \sigma \!\left(W_1 [X_t[i] \| M_t[i]] + b_1\right) + b_2,
    \label{eq:message_update}
\end{align}
where $X_t^{(l)} \in \mathbb{R}^{N \times d}$ is the feature matrix of counties at layer $l$, $\|$ denotes concatenation, $\sigma(\cdot)$ is an activation function, and $M_t^{(l)}$ represents the messages aggregated from neighboring counties according to the dynamic adjacency $\hat{A}^{(t)}$. If two counties are connected either geographically or through learned mobility, message passing allows their representations to influence one another proportionally to the edge weights in $\hat{A}^{(t)}$.

Each node embedding is then updated through a residual connection followed by layer normalization:
\begin{align}
    X_t^{(l+1)} = \text{LayerNorm}\!\left(X_t^{(l)} + Y_t\right),
    \label{eq:residual_update}
\end{align}
where $\text{LayerNorm}(\cdot)$ is defined as
\begin{align}
    \text{LayerNorm}(z) = \frac{z - \mu(z)}{\sqrt{\sigma^2(z) + \epsilon}} \gamma + \beta,
    \label{eq:layernorm}
\end{align}
with $\mu(z)$ and $\sigma^2(z)$ denoting the mean and variance computed along the feature dimension $d$, 
and $\gamma, \beta \in \mathbb{R}^d$ as learnable scale and shift parameters. The small constant $\epsilon$ ensures numerical stability.

After processing $L = 2$ message-passing layers, the network produces spatially embeddings $X^{(L)} \in \mathbb{R}^{B \times T \times N \times d}$, which encode higher-order spatial dependencies that reflect how regional conditions jointly influence HIV dynamics.

\subsection{Prediction}

After completing temporal encoding, mobility graph construction, and spatial message passing, the model generates the predicted HIV diagnosis rate for each county based on the final spatiotemporal embeddings. Let $X_{t,i}^{(L)}\in \mathbb{R}^d$ denote the representation of county $i$ at time $t$ from the last MPNN layer, and $H_{t,i}\in\mathbb{R^d}$ denote its corresponding temporal embedding from the Transformer encoder. At the most recent observed time step $T$, we combine the spatial and temporal embeddings through element-wise addition:
\begin{align}
    Z_i = X_{T,i}^{(L)} + H_{T,i},
    \label{eq:fusion}
\end{align}
where the fused representation $Z_i$ captures both dynamic spatial interactions and temporal evolution patterns specific to each county. This vector is passed through a prediction head with 2-layer MLP to estimate the county-level HIV diagnosis rate: $\hat{y}_i = W_2 \sigma(W_1 Z_i + b_1) + b_2$, where $W_1\in\mathbb{R}^{d_h\times d}$, $W_2\in\mathbb{R}^{1\times d_h}$, and $b_1,b_2$ are learnable parameters, $\sigma(\cdot)$ denotes activation function, and $d_h$ is the hidden dimension of the prediction MLP. 

\subsection{Loss Function}

The model is trained to minimize the mean squared error (MSE) between predicted and observed HIV diagnosis rates:
\begin{align}
    L = \frac{1}{BN} \sum_{b=1}^B \sum_{i=1}^N (\hat{y}_{b,i} - y_{b,i}^{(T)})^2,
    \label{eq:loss}
\end{align}
where $B$ denotes the batch size and $N$ denotes the number of counties per graph, $y_{b,i}^{(T)}$ denotes the observed HIV diagnose for county $i$ in batch $b$ at the final year $T$.

\section{Results}
\subsection{Model Evaluation Metrics}
We evaluate model performance using four criteria: the Mean Squared Prediction Error (MSPE), the Predictive Model Choice Criterion (PMCC) of Gelfand and Ghosh (1998)~\cite{gelfand1998model}, the Continuous Ranked Probability Score (CRPS) of Gneiting and Raftery (2007)~\cite{gneiting2007strictly}, and the Empirical Coverage Probability (ECP). Let $y_{\text{obs}}$ denote the observed county-level HIV diagnosis rates from 2008–2021 used for model training, and let $y_k$, $k=1,2,\dots,p$, denote the observed HIV diagnosis rates in 2022 for the $p$ test counties ($p=47$, $33$, and $112$ for Florida, California, and the New England states, respectively). $\hat{y}_k$ represents the point prediction generated by the MAT-MPNN model for county $k$, and $Y_k$ denotes the predictive distribution obtained from stochastic forward passes with Monte Carlo dropout~\cite{gal2016dropout} and ensemble variance estimation~\cite{lakshminarayanan2017deep}.

MSPE are defined as $\text{MSPE}=\frac{1}{p}\sum_{k=1}^{p}(\hat{y}_k-y_k)^2$, which measures the average squared deviation between predicted and observed values to quantify overall predictive accuracy. Lower MSPE values indicate higher accuracy.

Following Gelfand and Ghosh (1998), the Predictive Model Choice Criterion (PMCC) is computed as:

\begin{align}
    \text{PMCC}=\sum_{k=1}^p\frac{(\hat{y}_k-y_k)^2}{\text{Var}(Y_k)}+\sum_{k=1}^p\log\{\text{Var}(Y_k)\},
    \label{eq:PMCC}
\end{align}
where $\text{Var}(Y_k)$ represents the predictive variance for county $k$, estimated from multiple stochastic predictions of MAT-MPNN. The first term measures goodness of fit, while the second penalizes excessive predictive uncertainty. A smaller PMCC value indicates better predictive calibration.

The Continuous Ranked Probability Score (CRPS) is defined as:
\begin{align}
    \text{CRPS}=\frac{1}{p}\sum_{k=1}^{p}\Big(\mathbb{E}_{F_k}|Y_k - y_k| - \frac{1}{2}\mathbb{E}_{F_k}|Y_k - Y_k'|\Big),
    \label{eq:CRPS}
\end{align}
where $Y_k$ and $Y_k^{'}$ are independent samples from the predictive distribution $F$ for county $k$. CRPS is computed under the assumption that $Y_k$ follows a Gaussian distribution with mean $\hat{y}_k$ and variance $\text{Var}(Y_k)$. Lower CRPS values indicate sharper and better-calibrated predictive distributions.

Finally, the Empirical Coverage Probability (ECP) is defined as $\text{ECP}=\frac{1}{p}\sum_{k=1}^{p}\mathrm{1}(y_k\in [L_k^{(0.025)},U_k^{(0.975)}])$, where $[L_k^{(0.025)},U_k^{(0.975)}]$ denotes the $95\%$ predictive interval for county $k$, computed from the predictive mean and standard deviation of MAT-MPNN. ECP evaluates the calibration of predictive uncertainty, with values close to 0.95 indicating well-quantified uncertainty.

\begin{figure}[H]
    \centering
    \includegraphics[width=\textwidth]{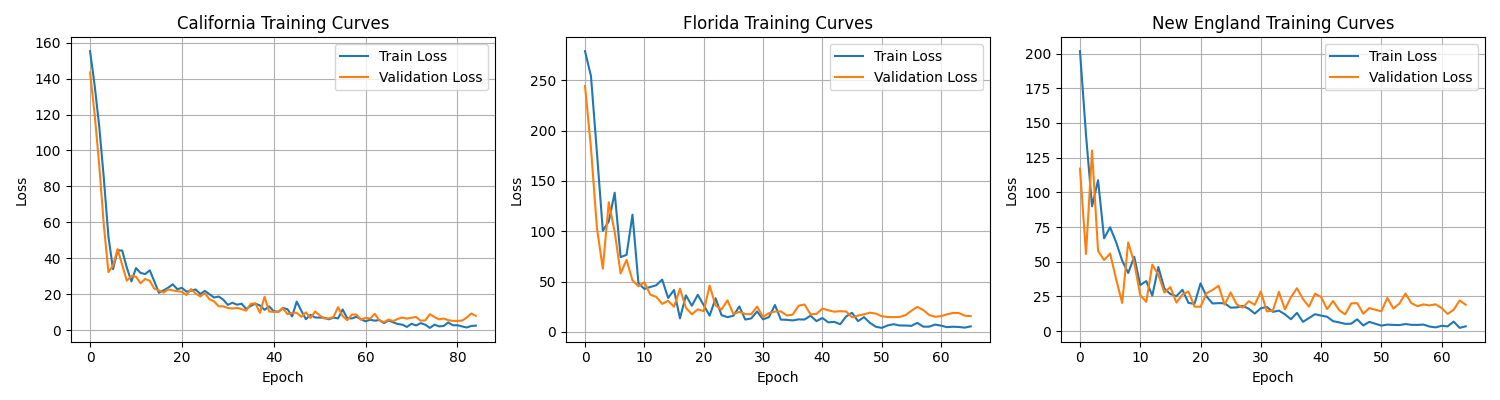}
    \caption{Training and validation loss curves of the MAT-MPNN for California (left), Florida (middle), and New England (right) across training epochs. The curves show smooth convergence and comparable train-validation behavior, indicating stable model fitting without severe overfitting.}
    \label{fig:training_curves}
\end{figure}

Fig.~\ref{fig:training_curves} presents the training and validation loss trajectories across epochs for the three regional models: California, Florida, and New England. Each curve represents the convergence behavior of the MAT-MPNN model during optimization. Training and validation losses decrease substantially in the early epochs and stabilize in the later epochs, implying that the model captured the underlying patterns without overfitting to the training data. The overall consistency between training and validation losses across regions demonstrates good generalization and stable model performance.

\subsection{Model Comparison}

To comprehensively assess the performance of MAT–MPNN, we compared it with representative temporal, spatial, and hybrid baselines. Temporal baselines (LSTM and Transformer) capture county‐level temporal dynamics independently. Spatial baselines (MPNN and GAT) use static county adjacency graphs to model spatial correlation but ignore temporal evolution. Hybrid baselines combine the two components (LSTM–MPNN, LSTM–GAT, Transformer–MPNN, and Transformer–GAT), sequentially modeling time and space through stacked architectures. Additionally, a statistical method was included for comparison: the Spatially Varying Auto-Regressive (SVAR) model of Shand et al.~\cite{shand2018svar}. All models were trained under identical data splits and optimization settings for fair evaluation.

\begin{table}[H]
    \centering
    \resizebox{\textwidth}{!}{%
    \begin{tabular}{lcccccccccccc}
        \toprule
        \multirow{2}{*}{\textbf{Model}} &
        \multicolumn{4}{c}{\textbf{Results for Florida}} &
        \multicolumn{4}{c}{\textbf{Results for California}} &
        \multicolumn{4}{c}{\textbf{Results for New England}} \\
        \cmidrule(lr){2-5} \cmidrule(lr){6-9} \cmidrule(lr){10-13}
        & MSPE & PMCC & CRPS & ECP & MSPE & PMCC & CRPS & ECP & MSPE & PMCC & CRPS & ECP \\
        \midrule
        LSTM & 48.52 & 180.02 & 3.97 & 0.829 & 16.50 & 114.78 & 2.29 & 0.900 & 19.60 & 279.38 & 2.49 & 0.879 \\
        Transformer & 55.76 & 192.32 & 3.26 & 0.932 & 15.53 & 104.91 & 2.34 & 0.618 & 12.78 & 356.64 & 2.06 & 0.787 \\
        MPNN & 104.30 & 224.21 & 4.48 & 0.955 & 31.26 & 148.81 & 3.01 & 0.941 & 31.41 & 435.88 & 3.04 & 0.980 \\
        GAT & 178.55 & 283.77 & 6.65 & 0.955 & 37.51 & 156.05 & 3.32 & 0.971 & 50.75 & 496.78 & 3.98 & 1.000 \\
        LSTM MPNN hybrid model & 74.51 & 206.57 & 5.11 & 0.771 & 21.73 & 126.43 & 2.58 & 0.867 & 21.13 & 282.48 & 2.57 & 0.864 \\
        LSTM GAT hybrid model& 84.04 & 280.53 & 4.21 & 0.886 & 15.40 & 160.14 & 2.30 & 0.794 & 15.24 & 366.51 & 2.05 & 0.901 \\
        Transformer MPNN hybrid model& 44.70 & 171.59 & 3.58 & 0.914 & 19.92 & 115.90 & 2.47 & 0.867 & 12.47 & 267.90 & 1.93 & 0.974 \\
        Transformer GAT hybrid model& 56.04 & 198.02 & 4.49 & 0.771 & 16.84 & 116.43 & 2.31 & 0.900 & 12.48 & \textbf{239.03} & 2.00 & 0.864 \\
        \textbf{MAT-MPNN} & \textbf{32.23} & \textbf{158.92} & \textbf{2.96} & 0.914 & 12.14 & 111.98 & 1.99 & 0.900 & \textbf{10.91} & 257.59 & \textbf{1.83} & 0.947 \\
        SVAR model Equation (3.2)\footnotemark & 75.30 & 221.80 & 3.91 & 0.957 & \textbf{7.23} & \textbf{94.84} & \textbf{1.46} & 0.970 & 45.88 & 393.70 & 2.56 & 0.920 \\
        \bottomrule
    \end{tabular}%
    }
    \caption{Model performance comparison across three regions. Lower MSPE, PMCC, and CRPS indicate better predictive performance.}
    \label{tab:regional_results}
\end{table}

Table \ref{tab:regional_results} summarizes the predictive performance of all models across the three study regions (Florida, California, and New England) using the four evaluation metrics (MSPE, PMCC, CRPS, and ECP). Overall, MAT–MPNN achieved the lowest MSPE and CRPS in Florida and New England, and performed competitively in California. Compared with Transformer–MPNN, MAT-MPNN reduces MSPE by approximately 27.9\% in Florida, 39.1\% in California, and 12.5\% in New England. The results show that combining dynamic mobility information with spatial–temporal learning significantly improves prediction accuracy, particularly in high mobility regions such as Florida.

From a calibration perspective, MAT–MPNN achieves smaller PMCC and CRPS values across all regions compared with the Transformer–MPNN baseline. Specifically, PMCC decreases from 171.59 to 158.92 in Florida, and from 115.90 to 111.98 in California, reflecting a better balance between fit and predictive uncertainty. In addition, CRPS decreases from 3.58 to 2.96 in Florida with a 17.3\% reduction, from 2.47 to 1.99 in California with a 19.4\% reduction, and from 1.93 to 1.83 in New England with a 5.2\% reduction. These improvements show that MAT–MPNN not only produces predictions that are closer to the true values but also gives more precise estimates of uncertainty. Furthermore, the ECP values of MAT–MPNN remain close to the nominal 0.95 level across all three regions, which also confirms that its predictive intervals are well calibrated.

Compared with the statistical SVAR model, MAT–MPNN performs better in Florida and New England, where HIV transmission patterns are more complex and higher mobility regions. In Florida, the MSPE decreases from 75.30 to 32.23 with a 57.2\% improvement, and in New England, it drops from 45.88 to 10.91 with a 76.2\% improvement. Although the SVAR has a slight advantage in California, its linear structure limits the performance in areas with high mobility patterns. These results show that MAT–MPNN captures both the time trends and spatial connections more effectively, providing a more accurate and reliable prediction framework for county-level HIV dynamics.

\footnotetext{SVAR model from Equation (3.2) in Shand et al.~(2018).}

\subsection{Visualization}

To provide an intuitive comparison between the baseline Transformer MPNN hybrid model and MAT-MPNN model, we visualized the map of predicted HIV diagnosis rates and absolute errors for the year 2022 (test set) across the three study regions: California, Florida, and New England. Each subfigure displays two rows of county-level maps: the top row indicates the result of Transformer MPNN hybrid model, and the bottom row shows the outputs from MAT-MPNN. The figures showed regions where the dynamic mobility-aware adjacency in MAT-MPNN reduces prediction error relative to the static adjacency used in Transformer MPNN hybrid model.

We focus visual analyses on Transformer MPNN hybrid model and MAT-MPNN because they share the same backbone architecture, where the only difference is in the adjacency matrix. This isolates the role of the MGG, allowing us to directly assess its effect on spatial prediction accuracy. The maps use a red color scale, where deeper shades of red indicate higher HIV diagnosis rates. Counties shaded in light gray represent areas with missing or suppressed data due to report limitations, consistent with the AIDSVu suppression rules.

\begin{figure}[H]
    \centering

    \begin{subfigure}[t]{0.48\textwidth}
        \centering
        \includegraphics[width=\textwidth]{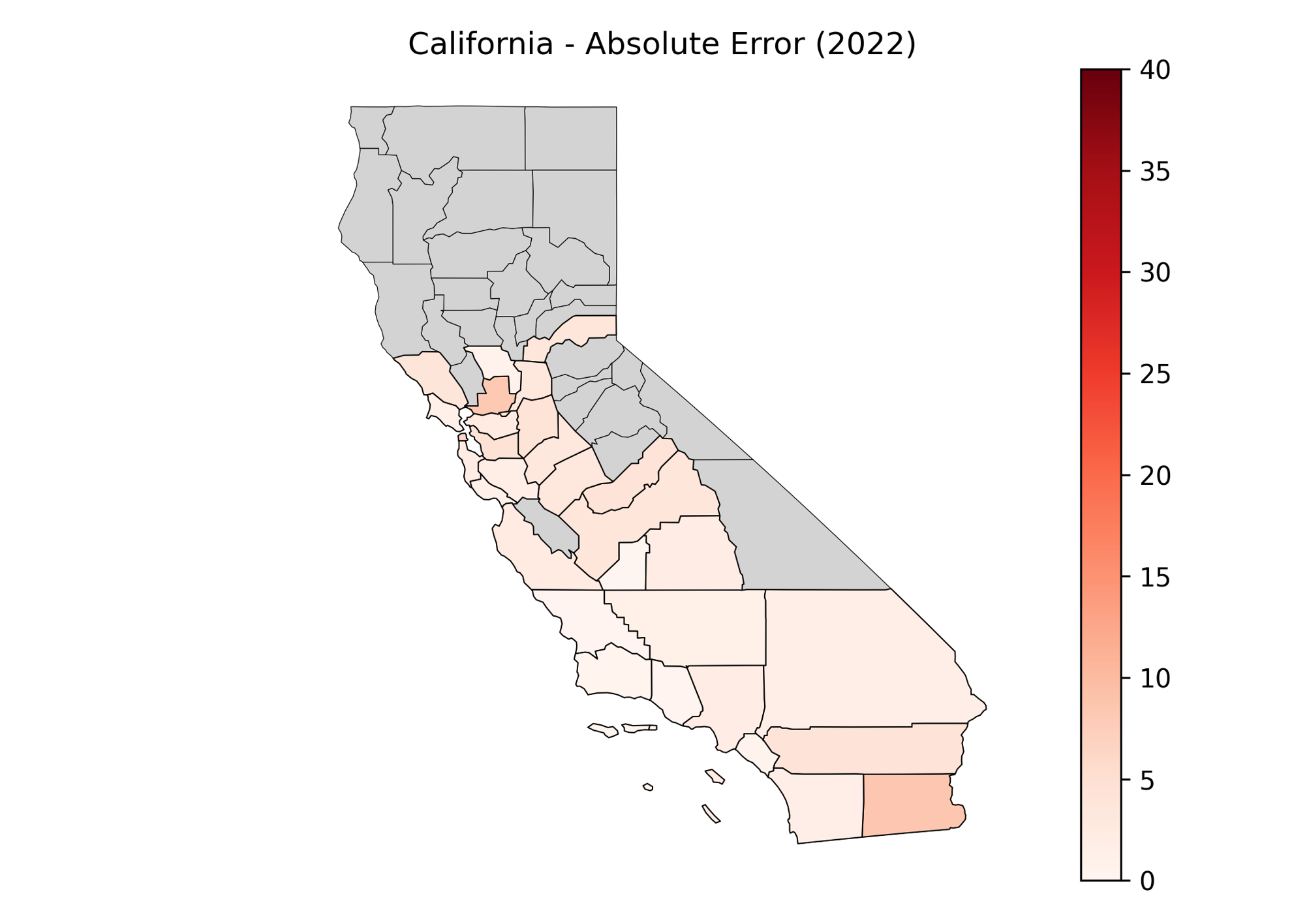}
    \end{subfigure}
    \hfill
    \begin{subfigure}[t]{0.48\textwidth}
        \centering
        \includegraphics[width=\textwidth]{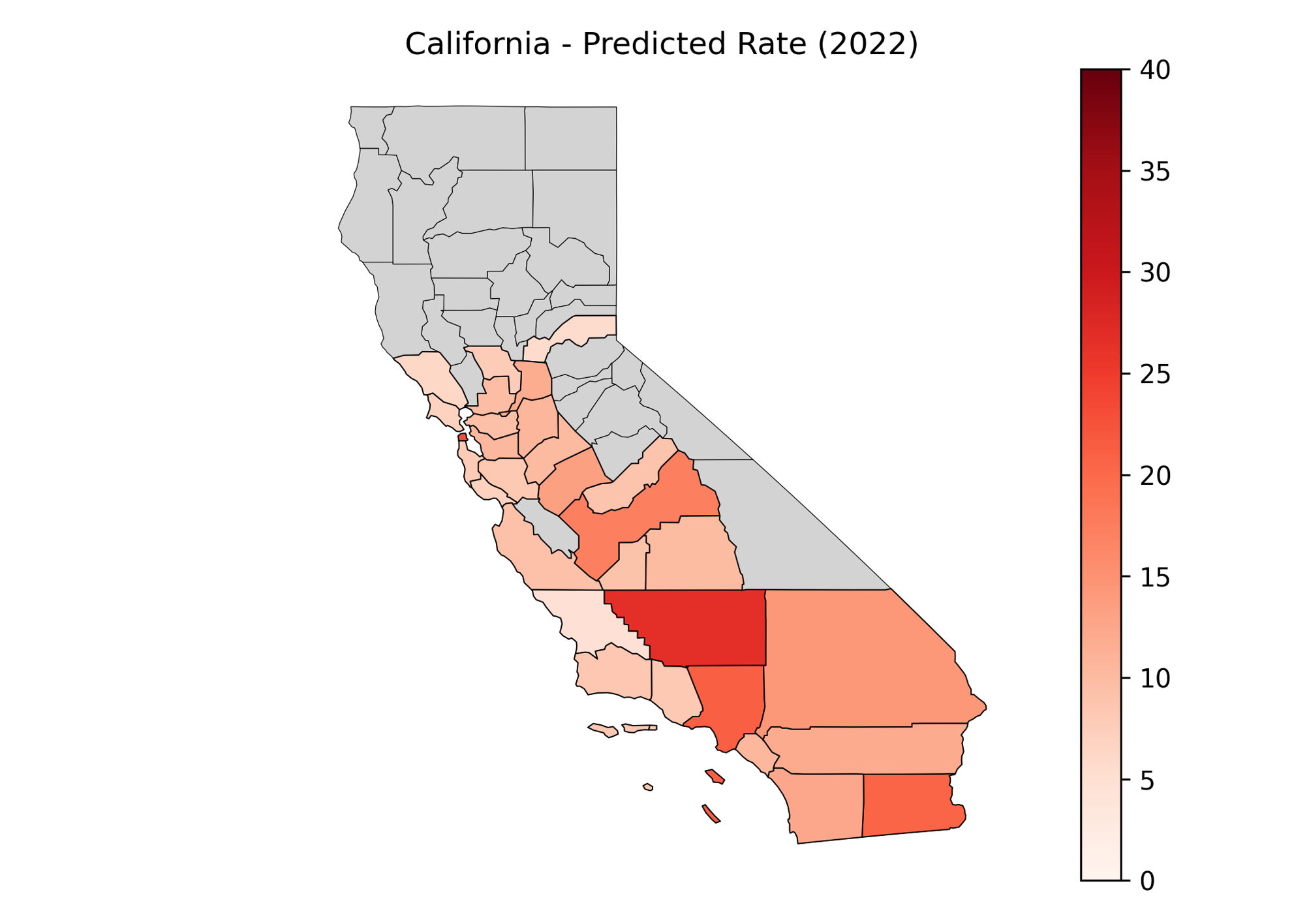}
    \end{subfigure}

    \begin{subfigure}[t]{0.48\textwidth}
        \centering
        \includegraphics[width=\textwidth]{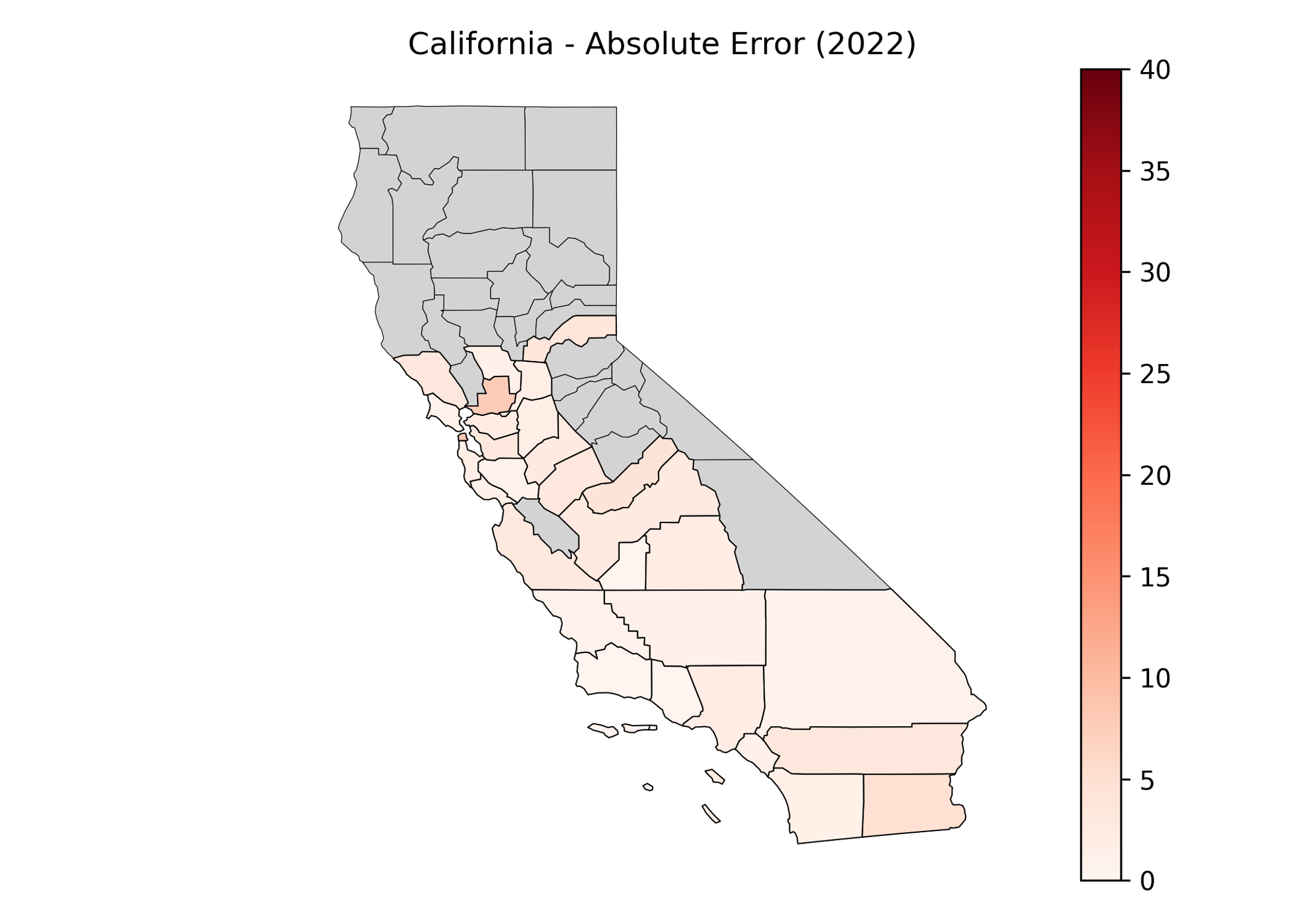}
    \end{subfigure}
    \hfill
    \begin{subfigure}[t]{0.48\textwidth}
        \centering
        \includegraphics[width=\textwidth]{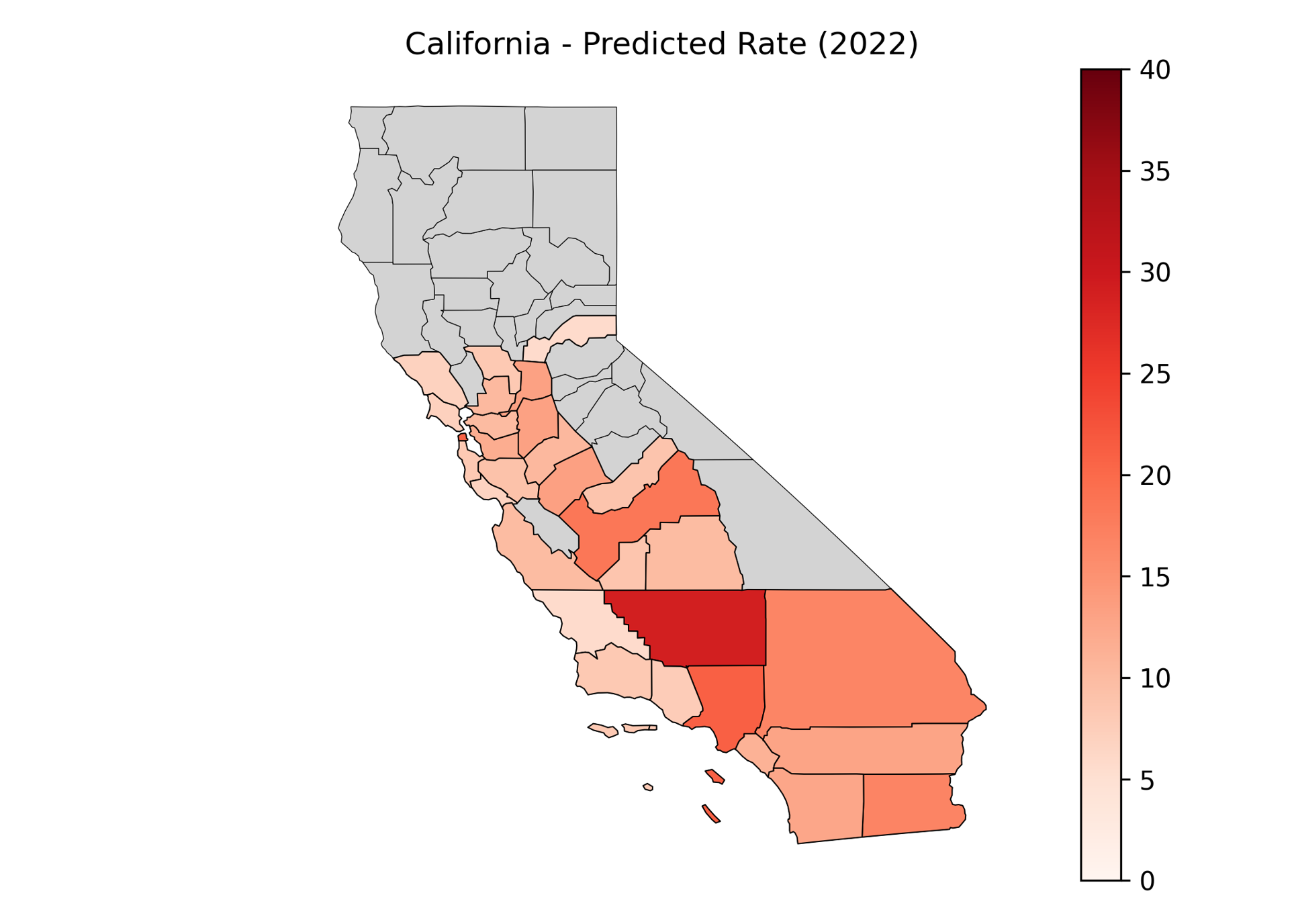}
    \end{subfigure}

    \caption{Comparison between Transformer MPNN hybrid model (top row) and MAT–MPNN (bottom row) predictions for California in 2022. The maps show absolute error (left) and predicted HIV diagnosis rates (right).}
    \label{fig:resultCA}
\end{figure}

\begin{figure}[H]
    \centering

    \begin{subfigure}[t]{0.48\textwidth}
        \centering
        \includegraphics[width=\textwidth]{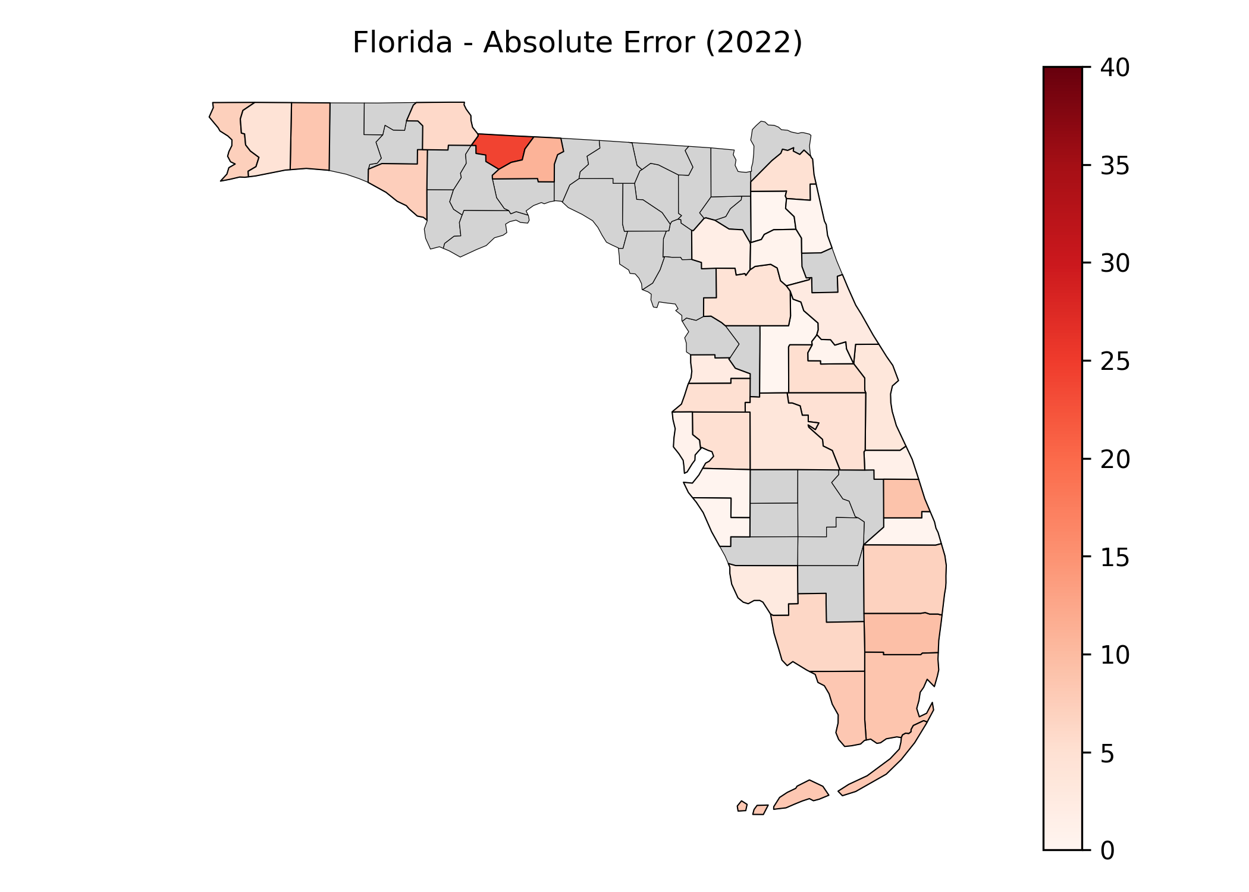}
    \end{subfigure}
    \hfill
    \begin{subfigure}[t]{0.48\textwidth}
        \centering
        \includegraphics[width=\textwidth]{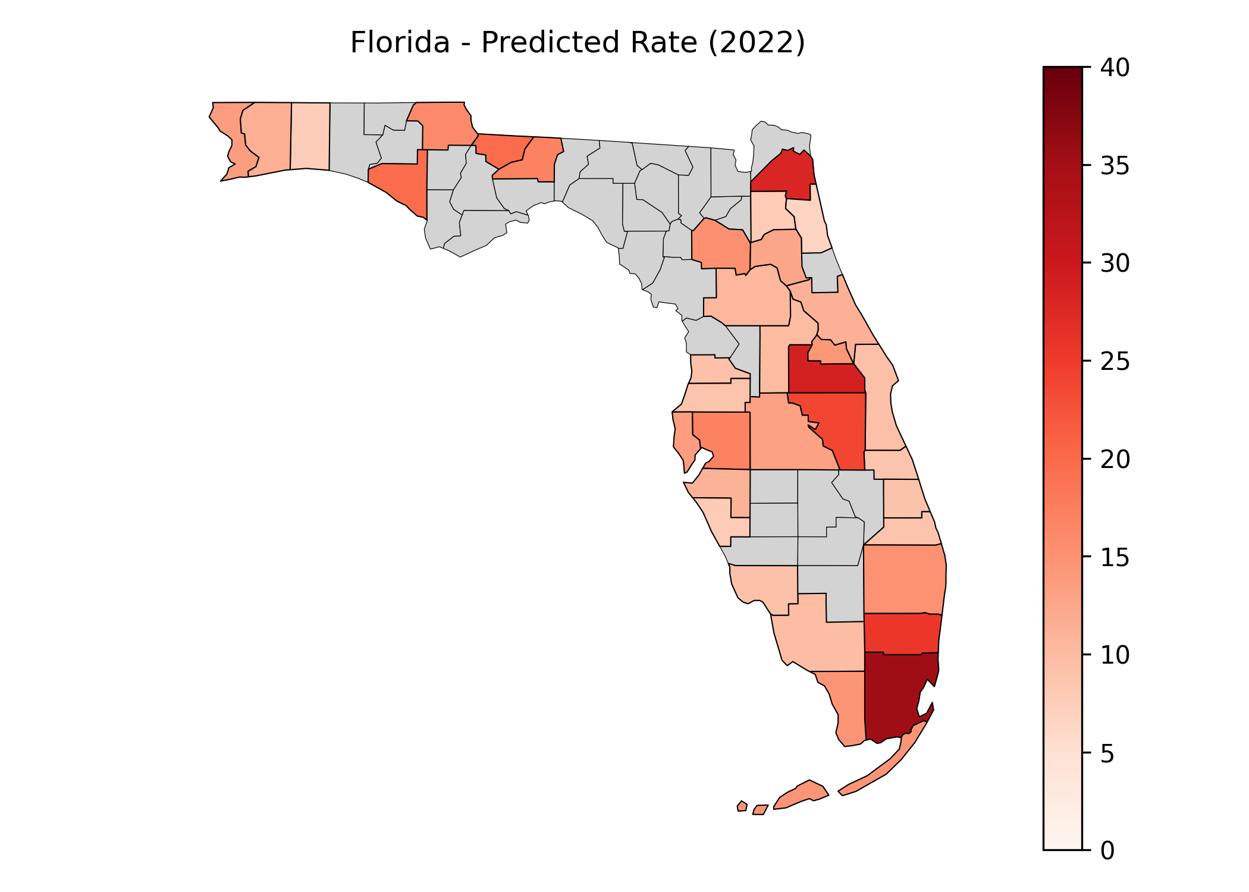}
    \end{subfigure}

    \begin{subfigure}[t]{0.48\textwidth}
        \centering
        \includegraphics[width=\textwidth]{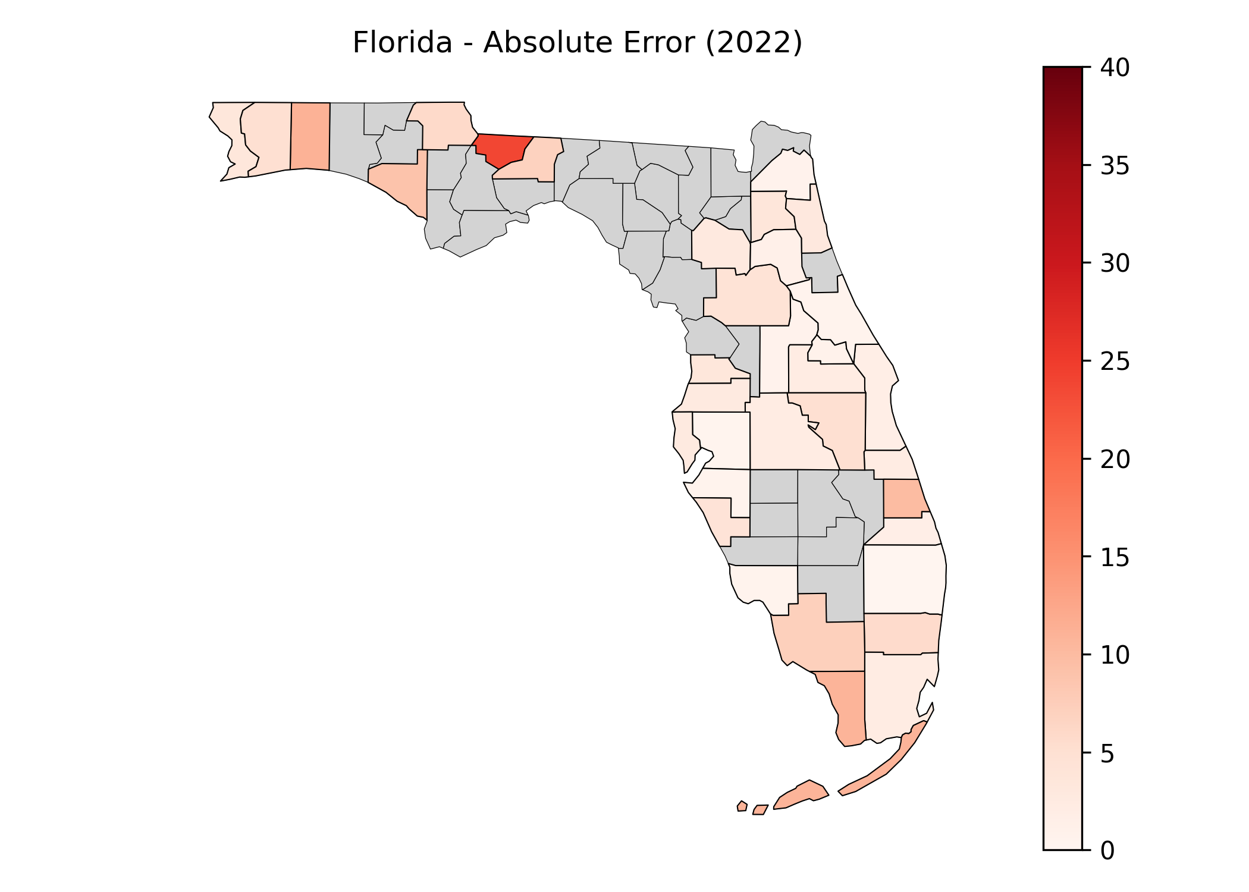}
    \end{subfigure}
    \hfill
    \begin{subfigure}[t]{0.48\textwidth}
        \centering
        \includegraphics[width=\textwidth]{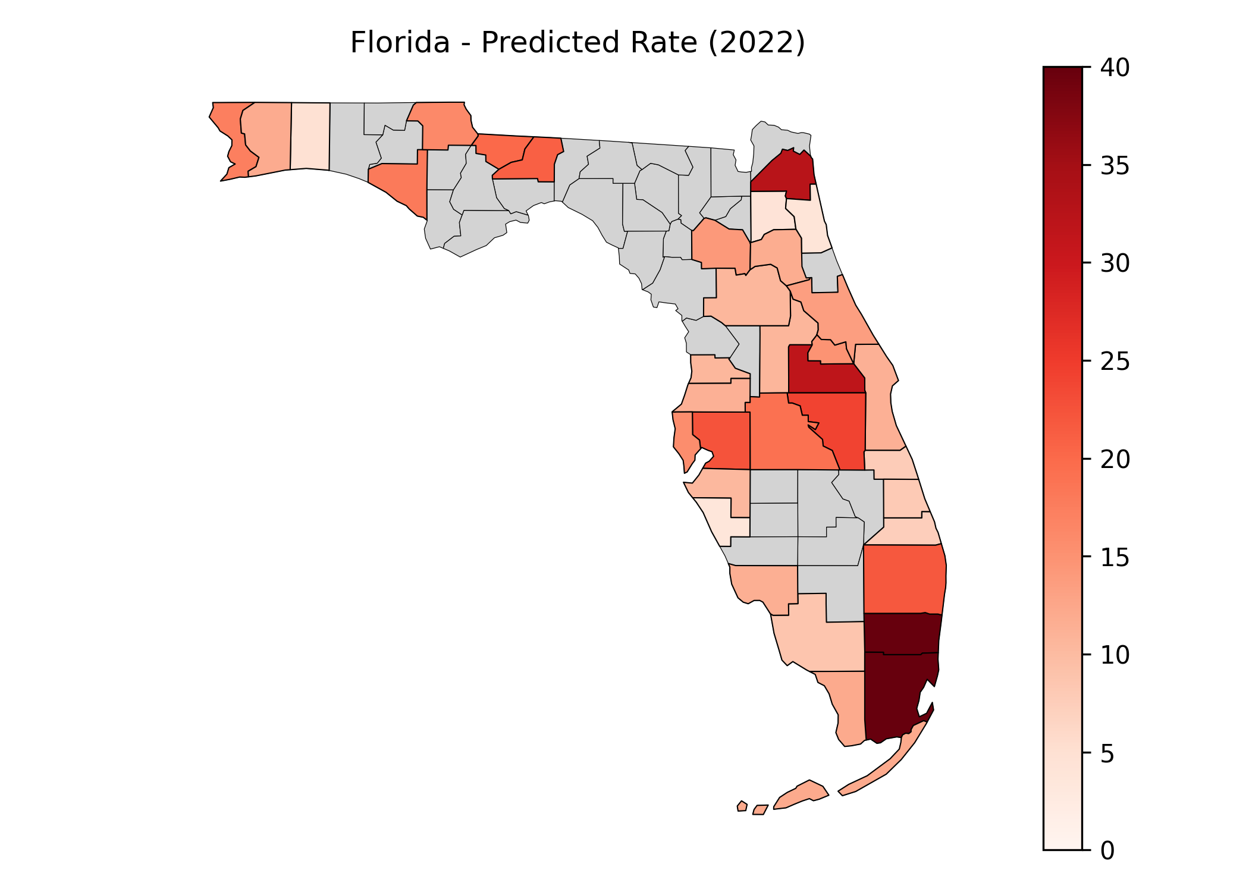}
    \end{subfigure}

    \caption{Comparison between Transformer MPNN hybrid model (top row) and MAT–MPNN (bottom row) predictions for Florida in 2022. The maps show absolute error (left) and predicted HIV diagnosis rates (right).}
    \label{fig:resultFL}
\end{figure}

\begin{figure}[H]
    \centering

    \begin{subfigure}[t]{0.48\textwidth}
        \centering
        \includegraphics[width=\textwidth]{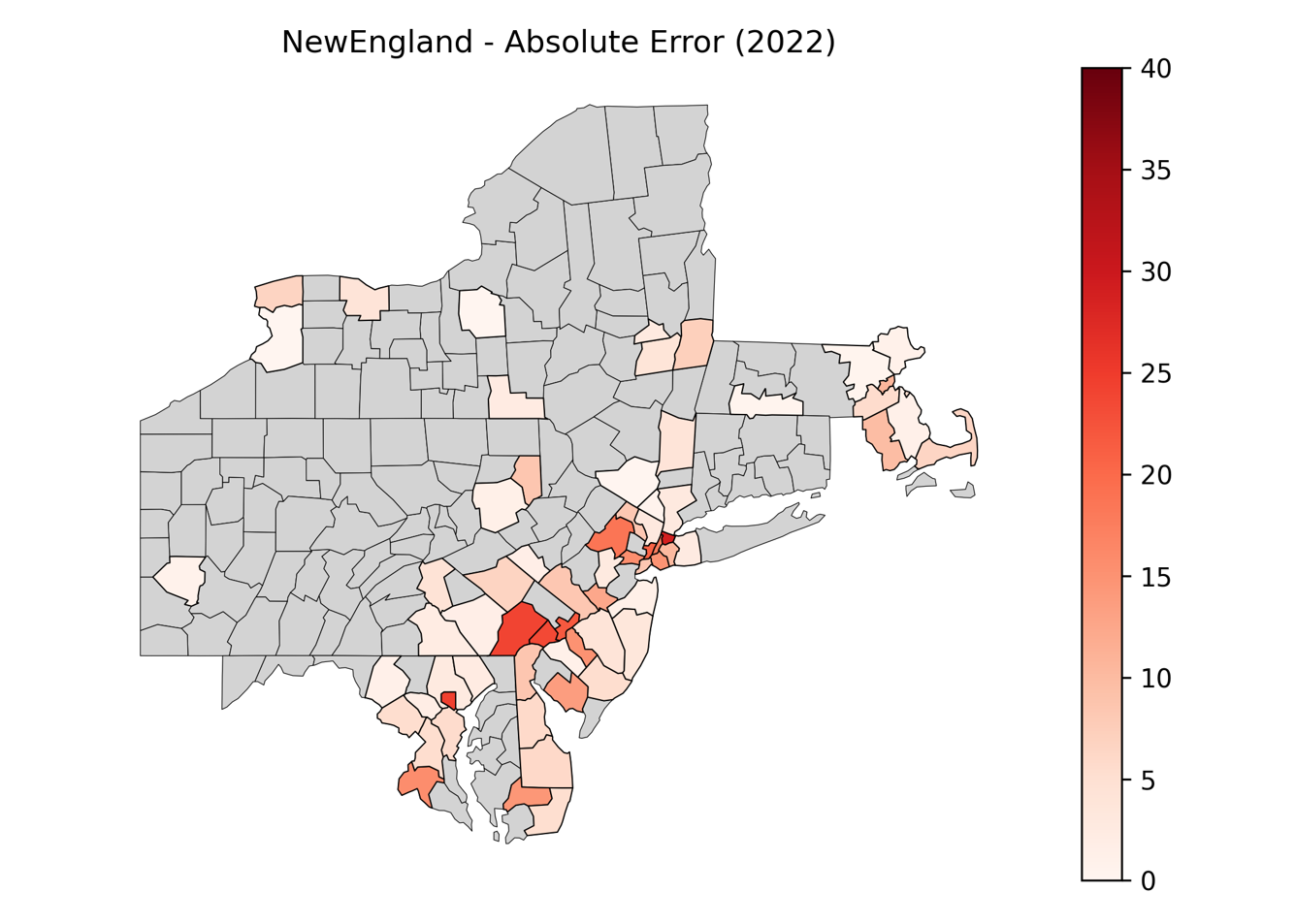}
    \end{subfigure}
    \hfill
    \begin{subfigure}[t]{0.48\textwidth}
        \centering
        \includegraphics[width=\textwidth]{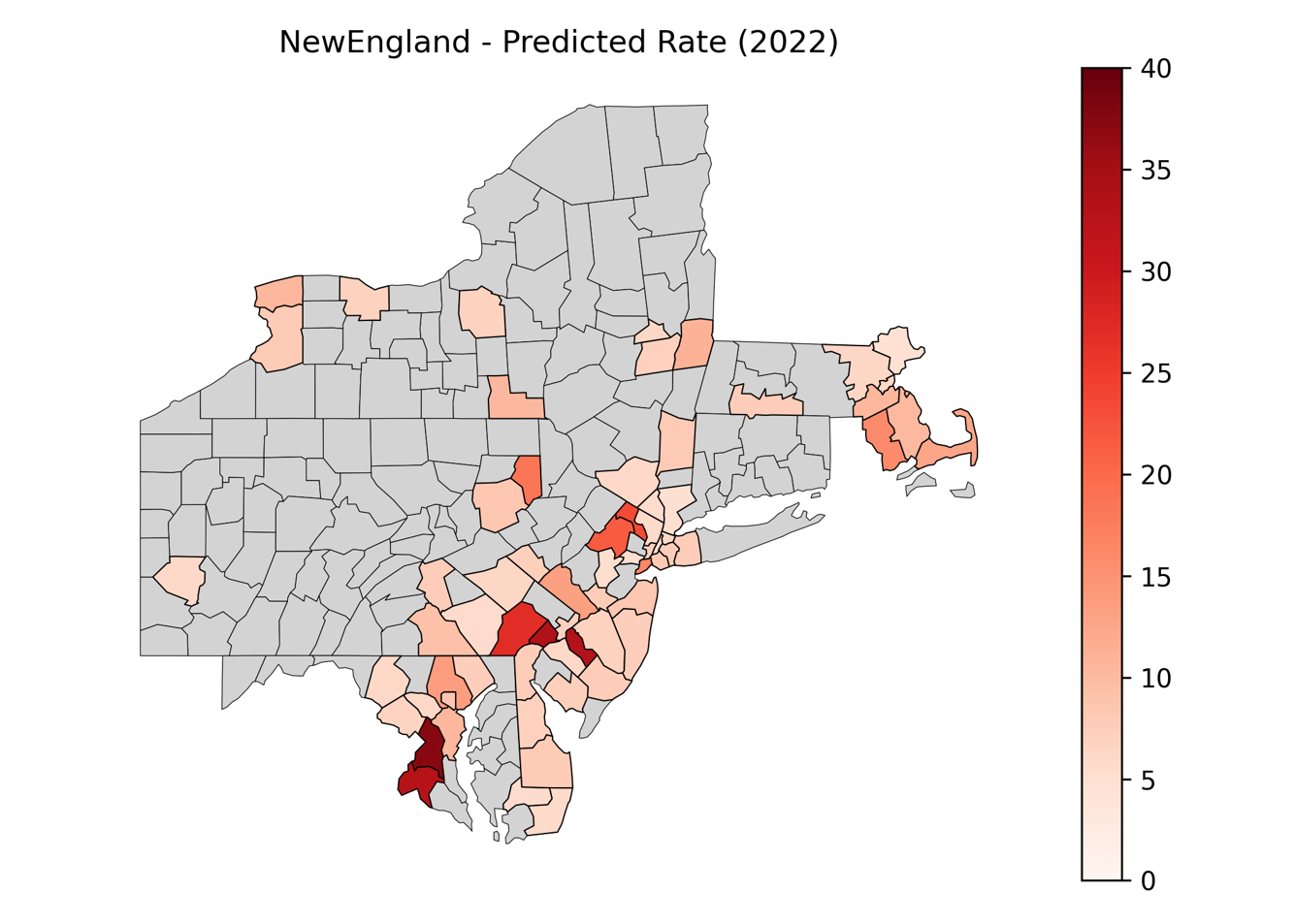}
    \end{subfigure}

    \begin{subfigure}[t]{0.48\textwidth}
        \centering
        \includegraphics[width=\textwidth]{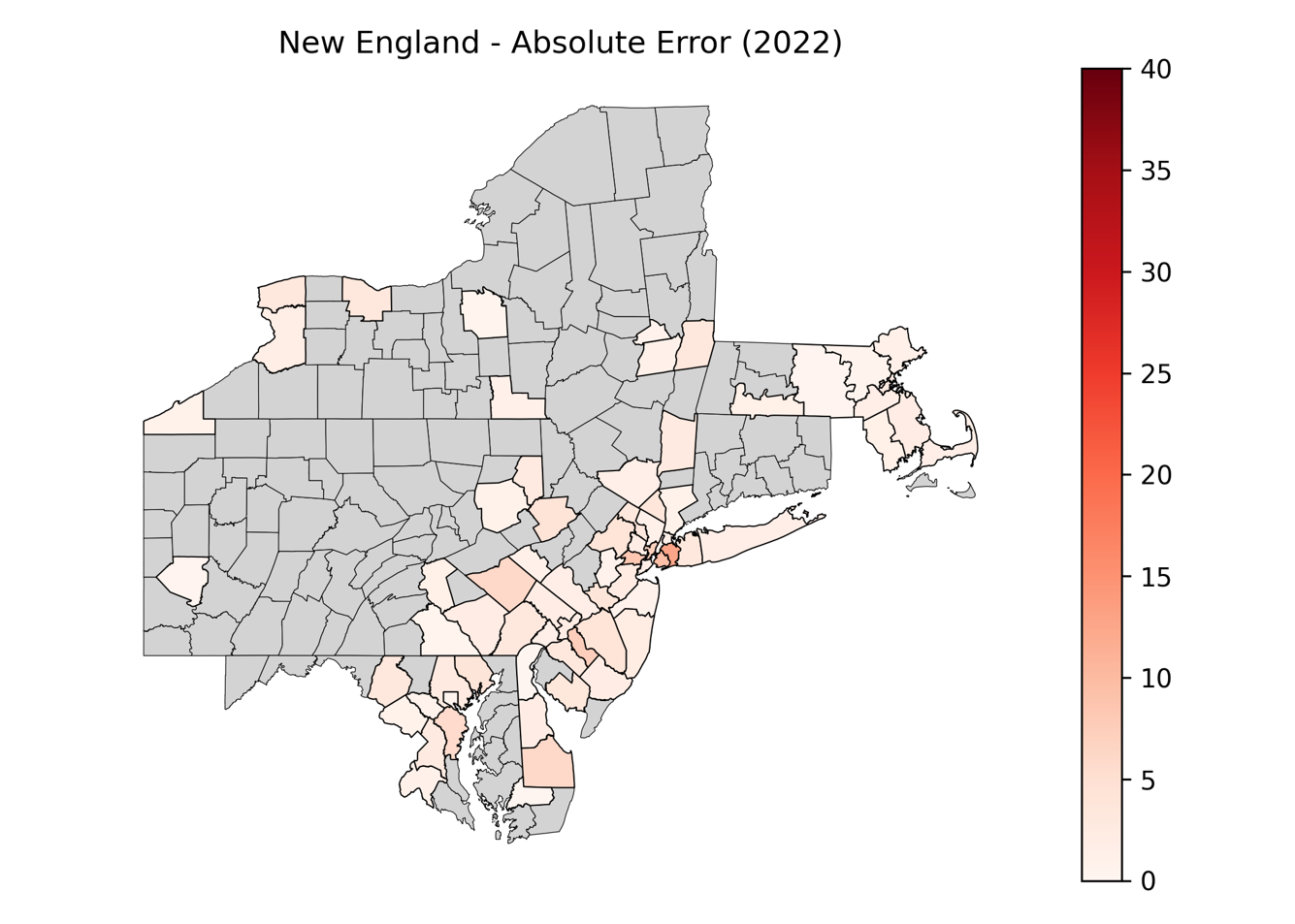}
    \end{subfigure}
    \hfill
    \begin{subfigure}[t]{0.48\textwidth}
        \centering
        \includegraphics[width=\textwidth]{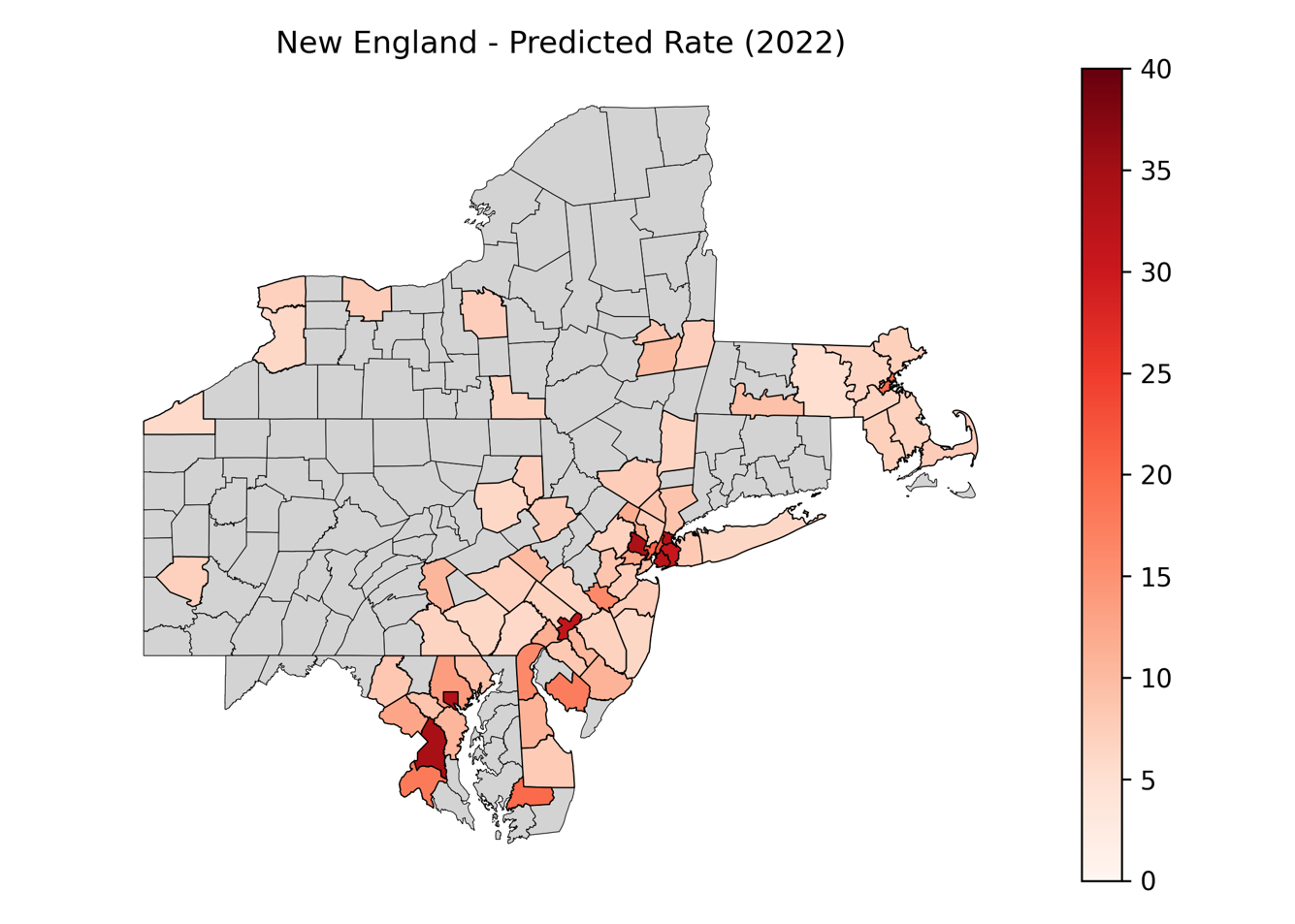}
    \end{subfigure}

    \caption{Comparison between Transformer MPNN hybrid model (top row) and MAT–MPNN (bottom row) predictions for New England region in 2022. The maps show absolute error (left) and predicted HIV diagnosis rates (right).}
    \label{fig:resultNE}
\end{figure}

As shown in Fig.~\ref{fig:resultCA}, Fig.~\ref{fig:resultFL}, and Fig.~\ref{fig:resultNE}, MAT-MPNN produces significantly lower absolute errors than Transformer MPNN hybrid model, particularly in large population density and high-mobility regions. This improvement meaning that mobility-aware dynamic adjacency effectively improves model sensitivity to real-world population movements and spatial transmission dynamics.

\subsection{Prediction for 2023}

\begin{figure}[H]
    \centering
    \begin{subfigure}[t]{0.45\textwidth}
        \centering
        \includegraphics[width=\textwidth, height=6cm]{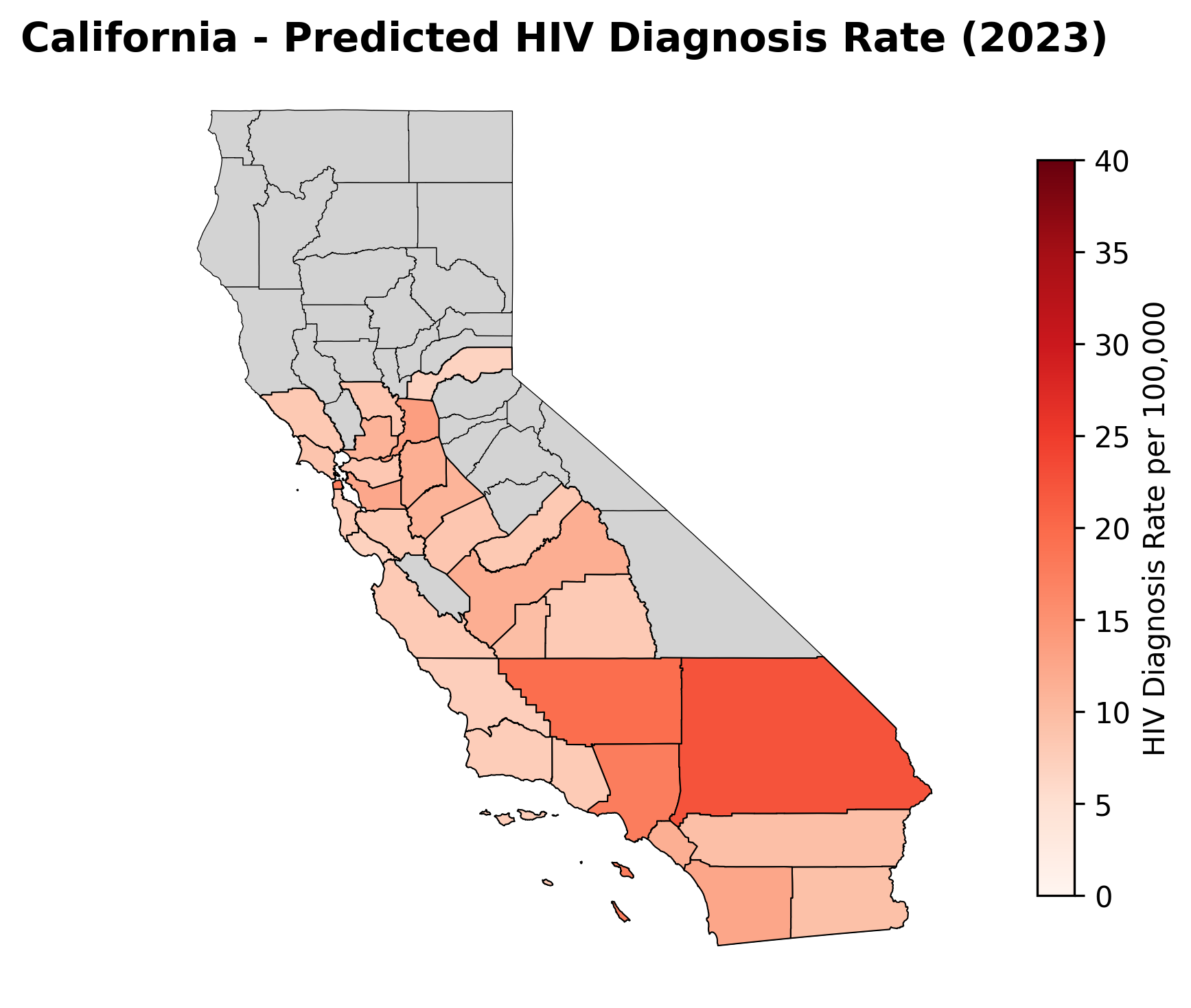}
        \caption{California}
    \end{subfigure}
    \hfill
    \begin{subfigure}[t]{0.45\textwidth}
        \centering
        \includegraphics[width=\textwidth, height=6cm]{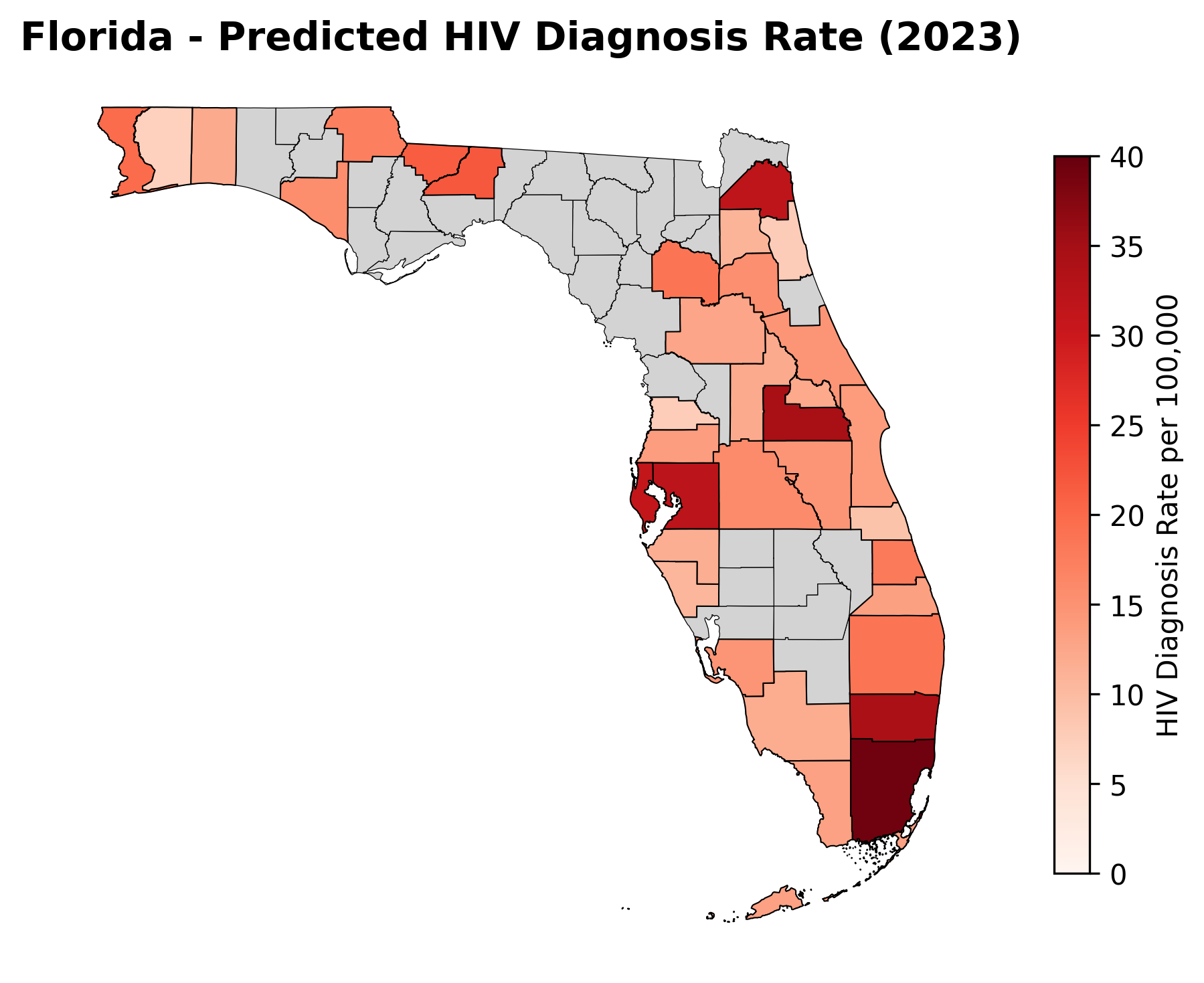}
        \caption{Florida}
    \end{subfigure}

    \vspace{3mm}
    \begin{subfigure}[t]{0.45\textwidth}
        \centering
        \includegraphics[width=\textwidth, height=6cm]{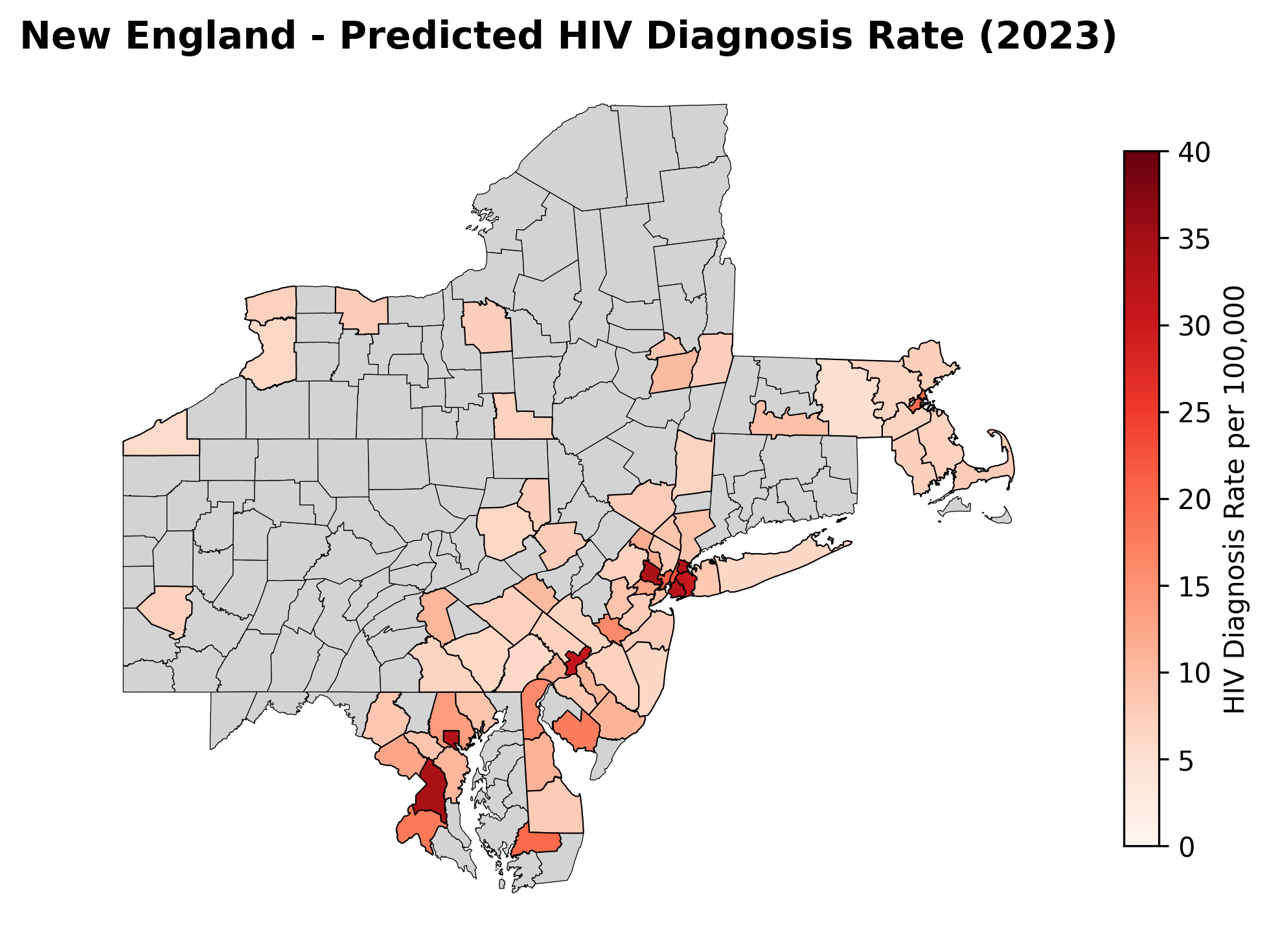}
        \caption{New England}
    \end{subfigure}

    \caption{Predicted HIV diagnosis rates for (a) California, (b) Florida, and (c) New England in 2023.}
    \label{fig:2023_pred}
\end{figure}

Using the trained MAT–MPNN model, we predicted county-level HIV diagnosis rates for 2023 across California, Florida, and the New England region. Fig.~\ref{fig:2023_pred} visualizes these predictions, where darker red shades represent higher estimated HIV rates per 100,000 population and light gray areas indicate missing or suppressed data.

Across all three regions, the predicted spatial patterns are consistent with known HIV diagnosis distributions. In California, higher diagnosis rates are concentrated in southern counties, particularly Los Angeles, San Diego, and Riverside, which have dense populations and higher transmission risk. In Florida, the model shows higher diagnosis rates in Miami, Broward, and Orange counties, which are also densely populated and known high-risk areas. In New England, high-risk areas appears along the southwestern corridor, mainly in Connecticut and Massachusetts, which are also densely populated and have higher mobility and infection risk.

\section{Discussion and Conclusion}

This study introduced a Mobility-Aware Transformer–MPNN (MAT–MPNN) framework for predicting county-level HIV diagnosis rates across three U.S. regions. Traditional methods such as autoregressive time-series models and LSTMs often assume fixed spatial structures and treat counties as independent units. GNNs like GCNs or GATs rely on static adjacency matrices based only on shared borders. Our framework overcomes these limitations by introducing a dynamic mobility-aware graph through the Mobility Graph Generator (MGG), which allows county relationships to change over time according to real world movement, economic links, and other regional interactions. Compared to baseline models, MAT–MPNN achieved lower MSPE, PMCC, and CRPS across all three regions, with the largest improvements in densely populated and highly connected areas. These results show that introducing mobility into the graph structure allows the model to capture how HIV spreads through human movement and social interaction.

The improvement in prediction accuracy comes from how MAT–MPNN learns both the when and the where of disease transmission. The Transformer captures long-term temporal patterns, while the MPNN uses dynamic graphs to understand how neighboring counties influence each other over time. This structure allows the model to learn evolving spatial dependencies that traditional static graphs cannot express. Besides HIV prediction, this framework produces a flexible way to model other infectious diseases that are influenced by mobility, environment, or social factors. In terms of methodology, it bridges epidemiology and modern deep learning, showing how mobility graphs can improve the interpretability of spatial prediction models. On a broader level, it also points out the importance of mobility information in public health and early intervention.

However, several limitations remain. Although MAT–MPNN integrates temporal mobility information, its performance may decline in regions with sparse data or low mobility coverage. Future work could focus on improving robustness in such settings and extending the model to cover all U.S. counties. In addition, county-level mobility data may not fully reflect within-county or short-distance movements that also contribute to transmission. Finally, due to data suppression and missing values in public HIV datasets, predictions for counties with limited information may carry a high uncertainty. Developing Bayesian extensions of MAT–MPNN could help quantify this uncertainty and improve interpretability.

In conclusion, MAT–MPNN provides a flexible and powerful framework for spatiotemporal disease modeling. By combining temporal encoding through the Transformer with dynamic graph message passing, it captures both temporal trends and evolving spatial relationships. Since dynamic human mobility plays a crucial role in disease transmission, future research could explore applications of this framework to other infectious diseases or health indicators.

\section*{Data Availability}

All datasets analyzed in this study are publicly available from official U.S. government and research institute sources.

County-level annual HIV diagnosis rates for 2008–2022 were obtained from the Centers for Disease Control and Prevention (CDC) AtlasPlus and the AIDSVu platform (\url{https://aidsvu.org/}), which disseminates data from the CDC’s National HIV Surveillance System (NHSS) in collaboration with Emory University’s Rollins School of Public Health.

Socioeconomic and demographic covariates were obtained from the U.S. Census Bureau’s \textit{American Community Survey (ACS) 5-Year Estimates}(\url{https://www.census.gov/programs-surveys/acs}), retrieved via the \textit{tidycensus} R package, and from the \textit{Small Area Health Insurance Estimates (SAHIE)} program (\url{https://www.census.gov/programs-surveys/sahie}).

County-level incarceration rate data were collected from the Vera Institute of Justice’s \textit{Incarceration Trends Dataset} (\url{https://github.com/vera-institute/incarceration-trends}).

Information on county hospital locations was obtained from the \textit{Homeland Infrastructure Foundation-Level Data (HIFLD)} (\url{https://hifld-geoplatform.opendata.arcgis.com/}).

Transportation infrastructure data, including Amtrak station locations, were obtained from the \textit{National Transportation Atlas Database (NTAD)} curated by the U.S. Bureau of Transportation Statistics (\url{https://data.bts.gov/}).

Geospatial boundary shapefiles for counties and states were obtained from the U.S. Census Bureau’s \textit{Cartographic Boundary Shapefiles (2022)} (\url{https://www.census.gov/geographies/mapping-files/time-series/geo/carto-boundary-file.html}).

All data were merged by Federal Information Processing Standards (FIPS) county codes (GEOID) and aligned by year. Some county-level HIV data are suppressed in accordance with CDC confidentiality policies and therefore cannot be redistributed.

\section*{Author Contributions}
\textbf{Conceptualization}: Zhaoxuan Wang, Weichen Kang, and Prof. Bo Li;\\
\\
\textbf{Methodology}: Zhaoxuan Wang, Yutian Han, and Lingyuan Zhao;\\
\\
\textbf{Software}: Zhaoxuan Wang and Weichen Kang;\\
\\
\textbf{Writing—original draft}: Zhaoxuan Wang, Yutian Han, and Lingyuan Zhao;\\
\\
\textbf{Writing—review $\&$ editing}: Zhaoxuan Wang, Weichen Kang, Yutian Han, Lingyuan Zhao, and Prof. Bo Li;\\
\\
\textbf{Supervision}: Prof. Bo Li.\\

\section*{Funding}
This research received no external funding.

\bibliographystyle{unsrt}  
\bibliography{references}

\end{document}